%% file: main.tex
\documentclass[a4paper,11pt]{article}
\usepackage{jheppub} %
\usepackage{lineno}
\usepackage{array,multirow,graphicx}
\usepackage{amssymb}
\usepackage{bbm}
\usepackage{hyperref}
\usepackage[dvipsnames]{xcolor}
\usepackage{float}
\usepackage{hhline}
\usepackage{multicol}
\usepackage{graphicx}
\usepackage{soul}
\usepackage{csquotes}
\usepackage{tikz-feynman}
\tikzfeynmanset{/tikzfeynman/every gluon@@/.style={
    /tikz/draw=none,
    /tikz/decoration={name=none},
    /tikz/postaction={
      /tikz/draw,
      /tikz/decoration={coil,aspect=0,
        amplitude=1.8mm,segment length=4.5mm, post length=0.5pt %
      },
      /tikz/decorate=true,
    }
}}
\usepackage{subcaption}
\usepackage{booktabs}
\usepackage{physics}
\usepackage{siunitx}
\usepackage{soul}
\RequirePackage[noabbrev]{cleveref}
\crefname{equation}{eq.}{eqs.}
\RequirePackage[status=draft,inline,nomargin,marginclue]{fixme}
\fxusetheme{color}
\FXRegisterAuthor{fh}{afh}{FH}
\FXRegisterAuthor{mc}{amc}{\color{purple}MC}
\FXRegisterAuthor{vg}{avg}{\color{orange}VG}

\newcommand{\alphaem}{\alpha_{\mathrm{em}}}
\newcommand*\diff{\mathop{}\!\mathrm{d}}

\title{VALO1.0: New real-photon parton distributions with Monte Carlo uncertainties}

\collaborationImg{\includegraphics{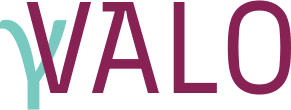}}

\author[a,b]{M. Chithirasreemadam,}
\author[a,b]{V. Guzey,}
\author[a,b]{F. Hekhorn,}
\author[a,b]{I. Helenius,}
\author[a,b]{H. Paukkunen}

\affiliation[a]{Department of Physics, University of Jyvaskyla, P.O. Box 35, FI-40014 University
of Jyvaskyla, Finland}
\affiliation[b]
{Helsinki Institute of Physics, University of Helsinki, P.O. Box 64, FI-00014 University of Helsinki, Finland}

\emailAdd{madhav.m.chithirasreemadam@jyu.fi}

\abstract{

Performing a global QCD analysis of data on the photon structure function $F_2^{\gamma}$ in $e^{+} e^{-}$ scattering, we determine new leading order (LO) and next-to-leading (NLO) parton distributions functions (PDFs) of the real photon. 
The resulting photon PDFs, referred to as VALO1.0, are obtained in the form of 
Monte Carlo (MC) replicas which assess the propagation of experimental uncertainties to the PDFs. 
To achieve well-converging fits, we employ a five-parameter hadron-like ansatz for the boundary conditions with simplifying assumptions on the flavor structure of the quark distributions and the large-$x$ behavior of the gluon distribution. 
This results in robust quark distributions at both LO and NLO and the gluon distribution at NLO with modest uncertainties, while leaving LO gluons still largely unconstrained.
The resulting photon PDFs broadly agree with the parameterizations available in the literature and set the stage for future analyses including additional 
photoproduction data, 
which could help to increase
the flexibility of
our input PDFs.

The LO and NLO VALO1.0 photon-PDF replicas, both in the DIS$_{\gamma}$ and $\overline{\rm MS}$ factorization schemes as well as the open-source $\gamma$\texttt{EKO} code for solving the scale dependence of photon PDFs and the analysis framework \texttt{VALOfitter} are made publicly available. 
}

\begin{document}
\maketitle
\flushbottom

\section{Introduction}
\label{sec:intro}

\input{sec_intro}
\section{Theory background to QCD analysis of photon PDFs}

\label{sec:theory}

\input{sec_theory}

\section{Fitting methodology}

\label{sec:meth}

\input{sec_meth}
\subsection{Overview of experimental data on $F_2^{\gamma}$}
\label{sec:data}
\input{sec_data}
\section{Results}
\label{sec:results}
\input{sec_results}

\section{Conclusions}
\label{sec:summary}

\input{sec_summary}

\acknowledgments

This research has been funded by the Center of Excellence in Quark Matter of the Research Council of Finland, projects 346325, 346326, and the Research Council of Finland projects 331545, 358090 and 361179. This research is associated with the Doctoral Education Pilot initiative for nuclear and particle physics of Ministry of Education and Culture.

\appendix

\section{Coefficient functions and splitting functions}
\label{app:splitting_functions}
\input{app_splitting_functions}
\section{$\gamma$\texttt{EKO}: point-like photon PDF evolution}
\label{app:geko}
\input{app_geko}

\bibliography{biblio}{}
\bibliographystyle{JHEP}

\end{document}

%% file: sec_intro.tex
In strong interactions the photon plays a dual role: it can interact with charged particles directly or through its fluctuations into hadronic states~\cite{Bauer:1977iq,Nisius:1999cv,Krawczyk:2000mf}. The corresponding photon-initiated reactions are referred to as the direct photon and the resolved photon processes, respectively. The resolved photon processes can be further divided into the point-like (anomalous) contribution, which corresponds to the photon splitting into a continuum of quark-antiquark pairs, and the hadronic contribution, where the photon can be viewed as a superposition of vector mesons.
The latter 
has traditionally been realized through the vector meson dominance (VMD) model~\cite{PhysRev.124.953, Sakurai1969, PhysRev.157.1376}. 
Note that these two contributions can be unified within 
the generalized vector meson dominance (GVMD) model~\cite{Kobayashi:1973ep,Fraas:1974gh}, where the photon state is expanded as an infinite series of vector mesons, whose parameters are adjusted to reproduce both
the total proton photoabsorption cross section and approximate scaling of
the $F_2^p$ structure function for low virtualities $Q^2$, see also the discussion in ref.~\cite{Schuler:1995fk}.

At high energies, when hadronic fluctuations of the photon
interact with a 
hard probe (e.g., with a virtual photon), one can resolve the underlying quark-gluon (parton) content of the photon.
As a result, in direct analogy with the 
familiar
proton 
or nucleus cases,
one can introduce parton distribution functions (PDFs) of the real photon $f_{j}^{\gamma}$
($j=q,g$ is the parton flavor) encoding the quark and gluon structure of the photon in Quantum Chromodynamics (QCD).
These PDFs are an essential non-perturbative component of QCD analysis of hard photon-induced processes within the framework of QCD factorization theorem~\cite{Collins:1989gx,CTEQ:1993hwr} and, thus, are as fundamental as proton and nucleus PDFs.

The principal source of information on $f_{j}^{\gamma}$ is electron-photon deep inelastic scattering (DIS) $e+\gamma \to e^{\prime}+X$, accessed through electron-positron scattering $e^{+}e^{-} \to e^{+}e^{-} \gamma^{\ast}  \gamma \to e^{+}e^{-}+ {\rm hadrons}$.
A series of measurements of the photon structure function $F_2^{\gamma}$ at CERN, DESY, KEK and SLAC stimulated developments in theoretical description of $F_2^{\gamma}$, first within the parton model~\cite{Walsh:1973mz,Kingsley:1973wk} and then in QCD~\cite{Witten:1977ju,Bardeen:1978hg,Gluck:1983bh,Moch:2001im}, and provided essential experimental input for global analyses of photon PDFs~\cite{Drees:1984cx,Abramowicz:1991yb,Gluck:1991jc,Gordon:1991tk,Hagiwara:1994ag,Schuler:1995fk,Gordon:1996pm,Gluck:1999ub,Cornet:2002iy,Cornet:2004nb,Aurenche:2005da,Slominski:2005bw}.
Additional constraints on $f_{j}^{\gamma}$ and tests of their universality are supplied by inclusive charged particle, dijet and charm production in photon-photon collisions~\cite{Nisius:1999cv,Krawczyk:2000mf} and by dijet, charm and beauty photoproduction in electron-proton scattering at  HERA~\cite{Butterworth:2005aq,Newman:2013ada}.

Global QCD analyses and resulting parameterizations of photon PDFs available in the literature~\cite{Drees:1984cx,Abramowicz:1991yb,Gluck:1991jc,Gordon:1991tk,Hagiwara:1994ag,Schuler:1995fk,Gordon:1996pm,Gluck:1999ub,Cornet:2002iy,Cornet:2004nb,Aurenche:2005da,Slominski:2005bw} address primarily the possibility of consistent description of $F_2^{\gamma}$ data using leading-order (LO) and next-to-leading order (NLO) perturbative QCD, employing initial conditions of various degrees of flexibility.
Additional constraints on the gluon distribution were examined by also including data on large-$p_T$ jet production in $\gamma \gamma$ scattering~\cite{Gordon:1996pm} and dijet production in $\gamma p$ scattering at HERA~\cite{Slominski:2005bw} (see also ref.~\cite{Aurenche:2005da}).
Several analyses also emphasized the charm quark distribution by making a comparison to the charm structure function $F_{2,c}^{\gamma}$ (charm production cross section) measured in $e^{+}e^{-}$ scattering~\cite{Hagiwara:1994ag,Cornet:2002iy,Cornet:2004nb}.
In addition, it was shown in ref.~\cite{Albino:2002ck} that QCD fits to $F_2^{\gamma}$ at large $x$ and $Q^2$ 
lead to a precise determination of the strong coupling constant $\alpha_s$. 

Recently, ultraperipheral collisions (UPCs) at the Large Hadron Collider (LHC)~\cite{Baltz:2007kq} have emerged as a new 
application and a potential 
probe
of photon PDFs, 
in particular, in 
inclusive~\cite{Guzey:2018dlm,Eskola:2024fhf,ATLAS:2024mvt} and diffractive~\cite{Guzey:2016tek} dijet photoproduction and inclusive open charm 
$D^0$~\cite{Cacciari:2025tgr,CMS:2025jjx} photoproduction in heavy ion UPCs.
In the future, measurements of dijets in quasi-real photoproduction in electron-proton scattering at the planned Electron-Ion Collider (EIC)~\cite{Guzey:2020zza} will be able to 
further 
constrain photon PDFs,
including their flavor separation~\cite{Chu:2017mnm}.

Addressing emerging demands for modern photon PDFs suited for phenomenological studies of UPCs at the LHC and photoproduction at the EIC~\cite{SURGE:2026hol}, in the present work, we 
obtain
new LO and NLO photon PDFs by performing a global QCD analysis of the world data on $F_2^{\gamma}$.
We employ hadron-like initial conditions for the quark and gluon distributions, which can be further adjusted and constrained by adding HERA and LHC photoproduction data.
Unlike most of the analyses available in the literature,
we quantify uncertainties of the resulting PDFs, for which we use Monte Carlo replicas.
We output and distribute our photon PDFs in the LHAPDF6 format, which should facilitate and standardize their phenomenological implementation, e.g., in general-purpose event generators for photoproduction~\cite{Helenius:2024rth,Hoeche:2023gme}. 
We appropriately name these new sets of PDFs as VALO, which is the Finnish word for ``light''.

The rest of this paper is organized as follows. 
\Cref{sec:theory} provides the necessary theoretical background to a QCD analysis of photon PDFs using the $F_2^{\gamma}$ structure function, where we also describe our procedure 
for scale evolution of photon PDFs, leaving details to \cref{app:splitting_functions,app:geko}.
In \cref{sec:meth}, we present the methodology of our analysis, including an overview of the fitting method based on Monte Carlo replicas, motivation of our choice for initial conditions for the photon PDFs, and a brief summary of the experimental data on $F_2^{\gamma}$
used in our fit.
\Cref{sec:results} contains the main results of this work, which include a detailed discussion of quality of our fits globally and differentially in different data sets,
the VALO1.0 LO and NLO photon PDFs at the initial scale and after scale evolution, comparison with selected earlier photon PDFs available in the literature, and a list of deliverables of this work.
We summarize our findings in \cref{sec:summary}.

%% file: sec_theory.tex
The main source of information on photon PDFs is electron-photon DIS $e(k)+\gamma(p) \to e^{\prime}(k^{\prime})+X$, see \cref{fig:dis_kin2}, where $k$ and $k^{\prime}$ are the momenta of the incoming and outgoing electrons, and 
$p$ and $q$ are the momenta of the real and virtual photon, respectively.

\begin{figure}[h]
\centering
\begin{tikzpicture}
   \begin{feynman}
        \vertex (mu);
        \vertex [right=2.5cm of mu] (f1);
        \vertex [right=1cm of f1] (f2);
        \vertex [right=2.5cm of f1] (f3);
        \vertex [blob, minimum size = 1.2cm, below=2.8cm of f2] (b) {};
        \vertex [above=0.7cm of f3] (nu);
        \vertex [below=0.5cm of b] (g);
        \vertex [left=3.55cm of g] (alpha);
        \vertex [right=2.6cm of b] (g1) {$ X$} ;
        \vertex [above=0.2cm of g1] (g2) {\phantom{\(X\)}};
        \vertex [below=0.2cm of g1] (g3) {\phantom{\(X\)}}; 
        \diagram* {
        (mu) -- [edge label'=$e(k)$] (f1) -- [edge label'=$e' (k')$] (nu),
        (f1) -- [gluon,  edge label=$\gamma^*({q= k-k'})$] (b),
        (alpha) -- [gluon, edge label'=$\gamma(p)$] (b), 
        (b) -- [] (g1),
        (b) -- [] (g2),
        (b) -- [] (g3),
        };
    \end{feynman}
\end{tikzpicture}
\caption{Electron-photon DIS $e+\gamma \to e^{\prime}+X$.}
\label{fig:dis_kin2}
\end{figure}

Similarly to DIS off a proton target, the corresponding DIS cross section can be expressed in terms of two structure functions
$F_2^{\gamma}(x,Q^2)$ and $F_L^{\gamma}(x,Q^2)=F_2^{\gamma}(x,Q^2)-2 xF_1^{\gamma}(x,Q^2)$,
\begin{equation}
\frac{d\sigma_{e\gamma \to e X}}{dx dQ^2} = 
\frac{2 \pi \alphaem^2}{x Q^4} \Big[F_2^{\gamma}(x,Q^2)(1+(1-y)^2)-y^2 F_L^{\gamma}(x,Q^2) \Big]
 \,,
\label{eq:dis_kin} 
 \end{equation}
where $\alphaem$ is the fine-structure constant, $y = (p\cdot q)/(p\cdot k)$ is the elasticity, 
$x=Q^2/[2 (p \cdot q)]$ is the Bjorken variable and $Q^2=-q^2$ is the virtuality of the probe photon. 
Since in electron-photon DIS experiments $y \ll  1$,
the contribution of $F_L^{\gamma}$ is commonly neglected. 

\subsection{The photon structure function $F_2^{\gamma}$}
\label{sec:coeff_fnc}

Theoretical description of the photon structure function $F_2^{\gamma}$ has been developing in parallel with that of the proton structure function $F_{2}^{p}$, revealing at the same time qualitative differences due to the possibility of the point-like coupling of photons to quarks. The first results on $F_2^{\gamma}$
were obtained in the parton model~\cite{Walsh:1973mz,Kingsley:1973wk}, which showed that $F_2^{\gamma}$ increases logarithmically with an increase of the photon virtuality $Q^2$ and 
that its $x$ dependence at large $Q^2$ can be rigorously predicted by calculating 
the $\gamma^{\ast} \gamma \to q {\bar q} \to \gamma^{\ast} \gamma$ box diagram, 
both in stark contrast with the proton case. 
Those were followed by analyses within pertubative QCD, which established the expressions for $F_2^{\gamma}$ to leading order (LO)~\cite{Witten:1977ju}, next-to-leading order (NLO)~\cite{Bardeen:1978hg,Gluck:1983bh} and next-to-next-to-leading order (NNLO)~\cite{Moch:2001im} accuracy.
They confirmed the positive logarithmic scaling violations of $F_2^{\gamma}$, which were observed in the parton model, and improved accuracy of QCD predictions for the shape of $F_2^{\gamma}$ (see, e.g., ref.~\cite{Buras:2005nj} for a discussion).

Our approach in 
the present work is based on the
QCD collinear factorization theorem~\cite{Collins:1989gx,CTEQ:1993hwr} and 
the formalism for the photon structure function $F_2^{\gamma}$ developed in refs.~\cite{Gluck:1983mm,Gluck:1991ee,Gluck:1991jc}. 
We  
limit ourselves to NLO accuracy because the main driver for phenomenological applications of our photon PDFs is photoproduction of di-jets at the LHC and EIC (see \cref{sec:intro}), whose numerical implementation is available only to this order~\cite{Guzey:2018dlm,Eskola:2024fhf,Guzey:2020zza}. 
Note, however, that there exist theoretical calculations of inclusive jet photoproduction with NNLO contributions~\cite{Klasen:2013cba}.

In the $\overline{\rm MS}$ factorization scheme and to NLO accuracy in the strong coupling $\alpha_s$ and to first non-vanishing order in $\alpha_{\rm em}$, the photon structure function $F_{2}^{\gamma}$ reads,
\begin{align}
\frac{1}{x} F_2^{\gamma}(x,Q^2) & = \sum_{j=1}^{n_f} e_{q_j}^2 \left(q_j^{\gamma}(x,Q^2)+{\bar q}_j^{\gamma}(x,Q^2)\right)\nonumber\\
&\hspace{20pt} + \frac{\alpha_s(Q^2)}{4 \pi}\sum_{j=1}^{n_f} e_{q_j}^2  \int^{1}_{x} \frac{\dd \xi}{\xi} C_q^{(1)}\left(\frac{x}{\xi}\right) \left(q_j^{\gamma}(\xi,Q^2)+{\bar q}_j^{\gamma}(\xi,Q^2)\right) \nonumber\\
&\hspace{20pt} + \frac{\alpha_s(Q^2)}{4 \pi} \sum_{j=1}^{n_f} e_{q_j}^2  \int^{1}_{x} \frac{\dd \xi}{\xi} C_g^{(1)}\left(\frac{x}{\xi}\right) g^{\gamma}(\xi,Q^2) + \frac{\alpha_{\rm em}}{4 \pi} \sum_{j=1}^{n_f} e_{q_j}^4 C_{\gamma}^{(1)}(x) \,,
\label{eq:F2_MSbar}
\end{align}
where $q_j^{\gamma}$, ${\bar q}_j^{\gamma}$, and $g^{\gamma}$ are the quark, anti-quark, and gluon distributions of the real photon, $e_{q_j}$ are quark electric charges, $j$ labels quark flavors, and $n_f$ is the number of active quark flavors.
Here and in the rest of the paper, we adopt the default scale choice by setting the renormalization scale $\mu_R$, introduced by the running of the strong coupling, and the factorization scale $\mu_F$, introduced by the collinear factorization, equal to the virtuality, $\mu_F^2 = Q^2 = \mu_R^2$.
In our analysis, we 
implement
the zero-mass variable flavor number scheme (ZM-VFNS)~\cite{Barontini:2024xgu} and change the number of light flavors $3 \leq n_f \leq 5$ depending on the value of $Q^2$.
We
regard the terms proportional to $\alpha_{\rm em}$ in \cref{eq:F2_MSbar} as an NLO correction and we do not consider higher order QED corrections 
to the hard scattering.
Thus, the functions $C_q^{(1)}$ and $C_g^{(1)}$ are the NLO quark and gluon coefficient functions, while $C_{\gamma}^{(1)}$ is the NLO photon coefficient function originating from the $\gamma \to q {\bar q}$ point-like splitting.
They are given by the standard expressions~\cite{Vermaseren:2005qc,Moch:2001im}, which we summarize in \cref{app:splitting_functions}.
For definiteness, we present below the coefficient function $C_{\gamma}^{(1)}$ specific to DIS on the real-photon target,
\begin{equation}
C_{\gamma}^{(1)}(z) = 4 N_c  \Bigg[\left(z^2+(1-z)^2\right) \ln\left(\frac{1-z}{z}\right)-1+8z(1-z) \Bigg] \,,
\label{eq:C1_gamma}    
\end{equation}
where $N_c=3$ is the number of colors.

In preparation for \cref{sec:evolution} addressing the scale evolution of photon PDFs, it is convenient to discuss the flavor dependence of photon PDFs in terms of the (hadronic) singlet component $\Sigma^\gamma$,
\begin{equation}\label{singlet}
    \Sigma^\gamma = \sum_{j=1}^{n_u} u_j^{\gamma,+} + \sum_{j=1}^{n_d} d_j^{\gamma,+}\,, \ \ \ \ \ q_j^{\gamma,+} = q_j^\gamma + \bar q_j^\gamma
\end{equation}
and the non-singlet component $\Sigma_\Delta^\gamma$~\cite{NNPDF:2024djq},
\begin{equation}
    \Sigma_\Delta^\gamma = \frac{n_d}{n_u} \sum_{j=1}^{n_u} u_j^{\gamma,+} - \sum_{j=1}^{n_d} d_j^{\gamma,+}\,,
\end{equation}
where $n_u$ and $n_d$ refer to the number of up-type (up and charm) and down-type (down, strange and bottom) quarks, respectively, and $n_f = n_u+n_d$.
Note that $\Sigma^\gamma$ and $\Sigma_\Delta^\gamma$ provide a minimal set of flavor combinations of the quark PDFs, which enter the description of the $F_2^{\gamma}$ structure function at NLO~\cite{NNPDF:2024djq}.
Defining the charge combinations
\begin{align}
    e_{\text{tot}}^2 = n_u e_u^2 + n_d e_d^2 \qand e^2_\Delta = e_u^2 - e_d^2 \,,
\end{align}
we can rewrite eq.~(\ref{eq:F2_MSbar}) in the following form:
\begin{align}
\frac{1}{x}F_2^{\gamma}(x,Q^2)&=\frac{n_u}{n_f}e^2_{\Delta}\qty(\left(\mathbbm 1 +\frac{\alpha_s(Q^2)}{4 \pi}C_q^{(1)} \right) \otimes \Sigma_\Delta^\gamma(Q^2))(x)\nonumber\\
&\hspace{20pt}+
\frac{e_{\text{tot}}^2}{n_f} \qty(\left(\mathbbm 1+\frac{\alpha_s(Q^2)}{4 \pi}C_q^{(1)}\right) \otimes \Sigma^{\gamma}(Q^2))(x)
\nonumber\\
&\hspace{20pt} + e_{\text{tot}}^2 \frac{\alpha_s(Q^2)}{4 \pi} \qty(C_g^{(1)} \otimes g^{\gamma}(Q^2))(x) + (n_u e_u^4 + n_d e_d^4) \frac{\alphaem}{4 \pi}C_{\gamma}(x) \,.
\label{eq:F2_MSbar_2}
\end{align}
In \cref{eq:F2_MSbar_2}, we used the standard notation for the convolution integrals involving the coefficient functions, e.g., for the quark non-singlet contribution,
\begin{align}
\qty(C_q^{(1)} \otimes \Sigma_{\Delta}^{\gamma}(Q^2))(x) & \equiv   \int^{1}_{x} \frac{d \xi}{\xi} C_q^{(1)}\left(\frac{x}{\xi}\right) \Sigma_{\Delta}^{\gamma}(\xi,Q^2) \,, \nonumber\\
\qty(\mathbbm 1 \otimes \Sigma_{\Delta}^{\gamma}(Q^2))(x) & \equiv  \int^{1}_{x} \frac{\dd \xi}{\xi} \delta\left(1-\frac{x}{\xi}\right) \Sigma_{\Delta}^{\gamma}(\xi,Q^2) =\Sigma_{\Delta}^{\gamma}(x,Q^2) \,,
\label{eq;convolution}
\end{align}
and similarly for the quark singlet and gluon terms.

To make the notation even more compact, we eventually adopt a vector notation, denoted in bold font henceforth, for photon PDFs, 
\begin{equation}
\vb f^{\gamma}(x,Q^2)=
\left(\begin{array}{c}
\Sigma_{\Delta}^{\gamma}(x,Q^2) \\
     \Sigma^{\gamma}(x,Q^2) \\
      g^{\gamma}(x,Q^2)
    \end{array}
 \right) \,,
 \label{eq:vector1}
\end{equation}
and for the coefficient functions,
\begin{align}
\vb C(z,\alpha_s(Q^2)) &= \vb C^{(0)}(z) + \frac{\alpha_s(Q^2)}{4\pi}\vb C^{(1)}(z) \nonumber\\
&=\left(\begin{array}{c}
     \frac{n_u}{n_f}e_\Delta^2 \delta(1-z)\\
     \frac{1}{n_f} e_{\text{tot}}^2 \delta(1-z)\\
      0
    \end{array}
 \right)^T + \frac{\alpha_s(Q^2)}{4\pi}\left(\begin{array}{c}
     \frac{n_u}{n_f}e_\Delta^2 C_q^{(1)} (z)\\
     \frac{1}{n_f} e_{\text{tot}}^2 C_q^{(1)} (z)\\
      e_{\text{tot}}^2 C_g^{(1)}(z)
    \end{array}
 \right)^T \,.
 \label{eq:vector2}
\end{align}
It allows us to obtain the following final expression for the photon structure function $F_2^{\gamma}$ in the $\overline{\rm MS}$ scheme at NLO accuracy:
\begin{equation}
\frac{1}{x}F_2^{\gamma}(x,Q^2)=\qty(\vb C(\alpha_s(Q^2)) \otimes \vb f^{\gamma}(Q^2))(x)
+ \qty(n_u e_u^4 + n_d e_d^4) \frac{\alphaem}{4 \pi}C_{\gamma}(x) \,.
\label{eq:F2_MSbar_3}
\end{equation}
To calculate $F_2^{\gamma}$ at LO in $\alpha_s$, one needs to set the NLO coefficient functions and the photon
coefficient function to zero, $\vb C^{(1)} = 0 = C_{\gamma}$, as well as to use the LO running for the strong coupling $\alpha_s$.

\begin{figure}[h]
\centering
\begin{subfigure}{0.5\textwidth}
\centering
\begin{tikzpicture}
   \begin{feynman}
        \vertex (mu);
        \vertex [right=2cm of mu] (f1);
        \vertex [below=0.5cm of f1] (f2);
        \vertex [above=0.5cm of f1] (f3);
        \vertex [above left=1.8cm of mu] (q) ;
        \vertex [below =1.5cm of mu] (alpha);
        \vertex [right=2 cm of alpha] (f2) ;
        \vertex [blob, minimum size = 0.3 cm, fill = none, below left= 1.4cm of mu](r){};
        \vertex [below left=0.9cm of mu] (l1);
        \vertex [right = 0.8cm of mu] (l2) ;
        \vertex [right = 0.4cm of r] (x) ;
        \vertex [above = 0.1cm of x] (r1) ;
        \vertex [below = 0.1cm of x]  (r2) ;
        \vertex [below left=2.8cm of mu] (p);
        \diagram* {
        (q) -- [photon, edge label'=\(\gamma^*\)] (mu) -- [fermion] (f1),
        (r) -- [fermion] (mu), 
        (p) -- [photon, edge label'=\(\gamma\)] (r),
        (r1) -- (r) -- (r2),
        };
    \end{feynman}
\end{tikzpicture} 
\caption{}
\label{subfig:resolved}
\end{subfigure}%
\begin{subfigure}{0.5\textwidth}
\centering
\begin{tikzpicture}
    \begin{feynman}
        \vertex (mu);
        \vertex [right=2cm of mu] (f1);
        \vertex [above left=1.8cm of mu] (q) ;
        \vertex [below =1.5cm of mu] (alpha);
        \vertex [right=2 cm of alpha] (f2) ;
        \vertex [below left=1.8cm of alpha] (p);
        \diagram* {
        (q) -- [photon, edge label'=\(\gamma^*\)] (mu) -- [fermion] (f1),
        (alpha) -- [fermion] (mu),
        (p) -- [photon, edge label'=\(\gamma\)](alpha),
        (alpha) -- [anti fermion](f2),
        };
    \end{feynman}
\end{tikzpicture}
\caption{}
\label{subfig:boxed}
\end{subfigure}%
\caption{The hadronic (a) and point-like (b) contributions to the photon structure function $F_2^{\gamma}$. }
    \label{fig:LOFeynman}
\end{figure}
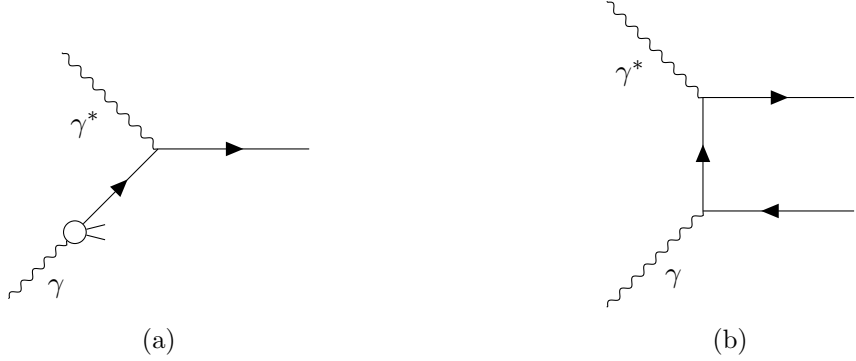

A schematic view of the two contributions in eq.~(\ref{eq:F2_MSbar_3}) is shown 
in figure~\ref{fig:LOFeynman}.
Graph (a) 
corresponds to the first term and represents the hadronic
contributions to $F_2^{\gamma}$ expressed in terms of photon PDFs, while 
graph (b) 
corresponds to the second term and represents the point-like contribution.

Another manifestation of the point-like coupling of the photon to quarks is the observation that beyond LO, the direct and resolved photon processes are 
connected by renormalization of photon PDFs and, hence, their separation is in general ambiguous and depends on the factorization scale and the factorization scheme.
The former can be readily seen from the scale evolution of photon PDFs written in the form explicitly including the photon-in-photon component, see, e.g., ref.~\cite{Klasen:2002xb}.

The factorization scale dependence of PDFs is typical for the QCD factorization framework.
In the case of photon PDFs, it is convenient to use the so-called DIS$_{\gamma}$ factorization scheme~\cite{Gluck:1991ee} (by analogy with the DIS scheme for the proton PDFs and 
$F_2^p$), which maximizes the resolved photon contribution by absorbing the entire point-like photon contribution to $F_2^{\gamma}$ into the definition of the quark photon PDFs.
In particular, one introduces the photon PDFs in DIS$_{\gamma}$ scheme by relating them to their counterparts in the $\overline{\rm MS}$ scheme,
\begin{align}
q_j^{\gamma}(x,Q^2)_{\rm DIS_{\gamma}} &= q^{\gamma}_j(x,Q^2)_{\overline{\rm MS}}
+e_{q_j}^2\, \frac{\alphaem}{8 \pi} C_{\gamma}(x) \,, \nonumber\\
{\bar q}_j^{\gamma}(x,Q^2)_{\rm DIS_{\gamma}} &= {\bar q}_j^{\gamma}(x,Q^2)_{\overline{\rm MS}}
+e_{q_j}^2\, \frac{\alphaem}{8 \pi} C_{\gamma}(x) \,, \nonumber\\
g^{\gamma}(x,Q^2)_{\rm DIS_{\gamma}} &= g^{\gamma}(x,Q^2)_{\overline{\rm MS}} \,.
\label{eq:scheme}
\end{align}
As a result, one 
arrives at
the following expression for the photon structure function $F_2^{\gamma}$ in the DIS$_{\gamma}$ scheme,
\begin{equation}
\frac{1}{x}F_2^{\gamma}(x,Q^2)=\qty(\vb C(\alpha_s(Q^2)) \otimes \vb f^{\gamma}(Q^2))(x) \,.
\label{eq:F2_DIS_gamma}
\end{equation}
In this equation, one implicitly assumes that the photon PDFs are evaluated in the DIS$_{\gamma}$ scheme.
Note that the scheme transformation affects photon PDFs only at the NLO accuracy, %
while at LO, where $C_\gamma = 0$, 
the two schemes are the same.
We
deliver our NLO PDFs in both $\text{DIS}_\gamma$  and $\overline{\text{MS}}$ schemes, where we obtain the latter by using the inverse of \cref{eq:scheme}.

In this work, we use the direct $x$-space representation and the DIS$_{\gamma}$ scheme for $F_2^{\gamma}$~(\ref{eq:F2_DIS_gamma}), which has several practical advantages for QCD fits.
First, it allows one to avoid numerical instabilities due to a singular behavior of  $C_{\gamma}(x)$ in the $x \to 1$ limit~\cite{Gluck:1991ee}.
Second, it makes the factorization formula for $F_2^{\gamma}$ identical to the proton case, which 
enables us
to use advanced fitting frameworks designed for DIS on the proton.
Notably, we use the Yadism package~\cite{Candido:2024rkr,barontini_2026_18758473} to compute the necessary quark and gluon coefficient functions in \cref{eq:F2_DIS_gamma}.
Moreover, this 
lets us
to use the full technology stack of the \texttt{pineline} framework~\cite{Barontini:2023vmr}, including fast interpolation grids provided by the \texttt{PineAPPL} library~\cite{Carrazza:2020gss,christopher_schwan_2025_15635174}.

\subsection{Scale evolution of photon PDFs}
\label{sec:evolution}
\input{sec_evolution}

\subsection{Generalized fast-kernel tables}
\label{sec:FK}

While we solve the scale evolution 
in 
Mellin space, our determination of
photon PDFs $\vb f^\gamma(x,\mu_F^2)$ using a global QCD analysis of the photon structure $F_2^{\gamma}$ is carried out in the $x$-representation. To this end, we design fast-kernel (FK) tables, allowing for efficient numerical calculations of $F_2^{\gamma}$.

As we have shown in \cref{sec:evolution}, the general solution 
to
scale evolution of photon PDFs can be written as a sum of the homogeneous and inhomegeneous components. Transforming \cref{eq:hominhom,eq:solution_hom} to the $x$-representation, one can write the photon PDFs $\vb f^\gamma(x,\mu_F^2)$ as
\begin{equation}
    \vb f^\gamma(x,\mu_F^2)  =(\vb E(\mu_F^2 \leftarrow \mu_0^2) \otimes \mathbf f^\gamma(\mu_0^2))(x) + \vb f^{\gamma}_{\rm inhom}(x,\mu_F^2) \,,
    \label{eq:pdfshort}
\end{equation}
where $\vb E$ is the hadronic EKO in the $x$-representation~\cite{Candido:2022tld}.
Similarly, \cref{eq:evolution_master} can also be written in $x$-space.

Since both the hadronic EKO $\vb E$ and the point-like contributions $\vb f^{\gamma}_{\rm inhom}$ are independent of the boundary condition $\vb f^\gamma(x,\mu_0^2)$, this allows us to extend the fast-kernel formalism~\cite{NNPDF:2014otw} developed 
within
the NNPDF framework to our case of photon PDFs.
Specifically, by inserting \cref{eq:pdfshort} in \cref{eq:F2_MSbar_3}, we can define both the hadronic and photonic FK tables by the following relations,
\begin{equation}
    \vb{FK}(Q^2,Q_0^2) = \vb C(\alpha_s(Q^2)) \otimes \vb E(Q^2\leftarrow Q_0^2) 
    \label{eq:FK1}
\end{equation}
and 
\begin{equation}
\mathrm{FK}^{\gamma}(x,Q^2) = (\vb C(\alpha_s(Q^2)) \otimes \vb f^\gamma_{\rm inhom}(Q^2))(x) \,,
\label{eq:FK2}
\end{equation}
which correspond to the (generalized) hadronic and point-like contributions, respectively.
Note that in \cref{eq:FK2} we use the DIS$_\gamma$ scheme and suppress a nontrivial dependence of $\vb{FK}(Q^2,Q_0^2)$ on the involved momentum fractions for brevity.

Using the FK tables, \cref{eq:F2_MSbar_3} can be rewritten in the following compact form,
\begin{equation}
    \frac 1 x F_2^{\gamma}(x,Q^2) = \qty(\vb {FK}(Q^2,Q_0^2) \otimes \vb f^{\gamma}(Q_0^2))(x) + \mathrm{FK}^{\gamma}(x,Q^2) \,.
    \label{eq:factFK}
\end{equation}
Equation~(\ref{eq:factFK}) is now optimal for fitting because all the perturbative ingredients can be pre-computed and the calculation of the photon structure function $F_2^{\gamma}(x,Q^2)$ at given $x$ and $Q^2$ in terms of the photon PDFs at the initial scale $\vb f^{\gamma}(x,Q_0^2)$ amounts to a fast linear algebra operation.

%% file: sec_evolution.tex
In the framework of collinear factorization of perturbative QCD, PDFs acquire a logarithmic dependence on the factorization scale $\mu_F^2$ (closely related to the virtuality  $Q^2$), which is a fundamental consequence of the renormalization (definition) of finite, physical PDFs.
In the case of photon PDFs, collinear divergences due to parton-parton and the real photon-quark splittings give rise to the scale evolution equations, which modify the familiar Dokshitzer-Gribov-Lipatov-Altarelli-Parisi (DGLAP) equations for proton PDFs by including an inhomogeneous term due to the 
$ \gamma \to q {\bar q}$ point-like contribution~\cite{Gluck:1983mm,Gluck:1991ee}. 

To illustrate our approach, we limit the discussion here to the case of PDF combinations in eq.~(\ref{eq:vector1}); the generalization including all flavours is straightforward \cite{NNPDF:2024djq}. Using the vector notation, eq.~(\ref{eq:vector1}), the factorization scale evolution equations
of photon PDFs in the $x$-space can be written in the following compact matrix form
\begin{equation}
\mu_F^2 \frac{\dd \vb f^{\gamma}(x,\mu_F^2)}{\dd \mu_F^2}=\vb k(x,\alpha_s(\mu_F^2))+(\vb P(\alpha_s(\mu_F^2))\otimes \vb f^{\gamma}(\mu_F^2))(x) \,,
\label{eq:ev_matrix}
\end{equation}
where $\vb P(\alpha_s(\mu_F^2))$ denotes the matrix of the parton-parton DGLAP splitting functions and $\vb k(x,\alpha_s(\mu_F^2))$ is the vector of the photon-parton splitting functions originating from the point-like coupling of the photon to quarks.
To NLO accuracy, the parton-parton splitting functions, $\vb P(\alpha_s(\mu_F^2))$, have the following form,
\begin{align}
\vb P(z,\alpha_s(\mu_F^2)) &= \frac{\alpha_{s}(\mu_F^2)}{4 \pi} \vb P^{(0)}(z)+\frac{\alpha_s^2(\mu_F^2)}{(4 \pi)^2} \vb P^{(1)}(z)
\nonumber\\
&=\frac{\alpha_{s}(\mu_F^2)}{4 \pi}\left(\begin{array}{ccc}
      P_{qq}^{(0)}(z) & 0 & 0\\
      0 & P_{qq}^{(0)}(z) &  (2 n_f) P_{qg}^{(0)}(z)  \\
      0 & P_{gq}^{(0)}(z) & P_{gg}^{(0)}(z) 
\end{array} \right) \nonumber\\
&\hspace{20pt}+\frac{\alpha_s^2(\mu_F^2)}{(4 \pi)^2} \left(
\begin{array}{ccc}
P_{ns+}^{(1)}(z) & 0 & 0\\
      0 & P_{qq}^{(1)}(z) & (2 n_f) P_{qg}^{1)}(z) \\
      0& P_{gq}^{(1)}(z) &  P_{gg}^{(1)}(z)
\end{array}
\right) \,,
\label{eq:P_expansion}
\end{align}
and the photon-parton splitting functions, $\vb k(x,\alpha_s(\mu_F^2))$, are given by
\begin{align}
\vb k(x,\alpha_s(\mu_F^2)) &= \frac{\alphaem}{4 \pi} \vb k^{(0)}(x)+\frac{\alphaem\alpha_s(\mu_F^2)}{(4 \pi)^2} \vb k^{(1)}(x) \nonumber\\
&=\frac{\alphaem}{4 \pi}\left(\begin{array}{c}
      k_{ns}^{(0)}(x) \\
      k_{\Sigma}^{(0)}(x) \\
      0
\end{array} \right) +\frac{\alphaem\alpha_s(\mu_F^2)}{(4 \pi)^2} \left(\begin{array}{c}
      k_{ns}^{(1)}(x) \\
      k_{\Sigma}^{(1)}(x)  \\
      k_{g}^{(1)}(x) 
\end{array} \right) \,.
\label{eq:k_expansion}
\end{align}
For illustration, we present their explicit form at LO,
\begin{align}
k_{ns}^{(0)}(x) &= 4 N_c n_d e_{\Delta}^2 \cdot (x^2+(1-x)^2) \,, \nonumber\\
k_{\Sigma}^{(0)}(x)&= 4 N_c  e^2_{\rm tot} \cdot (x^2+(1-x)^2) \,.
\label{eq:k_lo}
\end{align}
Note that the functional form of the $x$-dependence of these LO quark splitting functions
comes from a direct calculation of the logarithmic contribution to the quark box diagram, corresponding to graph (b) in \cref{fig:LOFeynman}.

One can see from \cref{eq:P_expansion} that the non-singlet quark combination $\Sigma^{\gamma}_{\Delta}$ evolves autonomously, while the quark singlet component $\Sigma^{\gamma}$ mixes with gluons under scale evolution, in direct analogy with the proton case.
Also, as follows from eq.~(\ref{eq:k_expansion}), the inhomogeneous term for gluons is absent at LO since the photon does not couple directly to gluons.
Finally, we note that the factorization scheme transformation, \cref{eq:scheme}, which we discussed in \cref{sec:coeff_fnc}, also affects the quark and gluon inhomogeneous terms $k_{ns}^{(1)}(x)$, $k_{\Sigma}^{(1)}(x)$, and $k_{g}^{(1)}(x)$, starting from NLO accuracy~\cite{Gluck:1991ee}. The splitting functions used in \cref{eq:P_expansion,eq:k_expansion}, both in
$\overline{\rm MS}$ and DIS$_{\gamma}$ schemes, are further discussed  in \cref{app:splitting_functions}.

In our analysis, we solve the scale evolution equations for photon PDFs~(\ref{eq:ev_matrix}) in Mellin moment space.
To this end, we adopt the notation and conventions of the \texttt{EKO} (Evolution Kernel Operator) framework~\cite{Candido:2022tld,barontini_2025_15878535} developed in the context of proton PDF fits 
within
the NNPDF collaboration.
For a detailed discussion of all available settings and options in \texttt{EKO}, we refer the reader to the dedicated paper~\cite{Candido:2022tld}, the source code~\cite{barontini_2025_15878535}, and the online documentation\footnote{\url{https://eko.readthedocs.io}}.

Performing a Mellin transformation of the vectors of photon PDFs, $\vb f^{\gamma}(x,\mu_F^2)$, the inhomogeneous splitting functions, $\vb k(x,\mu_F^2)$, and the matrix of the splitting functions, $\vb P(\alpha_s(\mu_F^2))$,
\begin{align}
\tilde{\mathbf{f}}^{\gamma}(N,\mu_F^2) & \equiv \int\limits_0^1 \! \diff x\, x^{N-1} \vb f^{\gamma}(x,\mu_F^2) 
\,, \nonumber\\
\tilde{\mathbf{k}}(N,\alpha_s(\mu_F^2)) & \equiv  -\int\limits_0^1 \! \diff x\, x^{N-1} \vb k(x,\alpha_s(\mu_F^2)) 
\,, \nonumber\\
\mathbf{\gamma}(N,\alpha_s(\mu_F^2)) & \equiv  -\int\limits_0^1 \! \diff x\, x^{N-1} \vb P(x,\alpha_s(\mu_F^2))  \,,
\label{eq:Mellin} 
\end{align}
the integro-differential evolution equations for photon PDFs $\tilde{\mathbf{f}}^\gamma$ in the $x$-representation, \cref{eq:ev_matrix}, become ordinary coupled differential equations in Mellin space~\cite{Moch:2001im},
\begin{equation}
\mu_F^2 \frac{\diff \tilde{\mathbf{f}}^{\gamma}(N,\mu_F^2)}{d\mu_F^2} = -\gamma(N,\alpha_s(\mu_F^2)) \tilde{\mathbf{f}}^{\gamma}(N,\mu_F^2) - \tilde{\mathbf k}(N,\alpha_s(\mu_F^2)) \,.
\label{eq:DGLAP}
\end{equation}
Note that in \cref{eq:Mellin,eq:DGLAP} we use the sign convention of ref.~\cite{Moch:2001im}. 
To enhance transparency of our following algebraic derivations, we will  
suppress the explicit dependence of all quantities in Mellin space on the Mellin moment $N$.

Applying the standard method for solving inhomogenous differential equations, we can express the solution of \cref{eq:DGLAP} as a sum of the general solution of the homogenous equations and a particular solution of the full inhomemogenous equation.
We denote those as an homogeneous (hadronic) component $\mathbf f^\gamma_{\rm hom}$ and an inhomogeneous (point-like) component $\mathbf f^{\gamma}_{\rm inhom}$, respectively,
\begin{align}
    \tilde{\mathbf f}^\gamma(\mu_F^2) = \tilde{\mathbf f}^\gamma_{\rm hom}(\mu_F^2)  + \tilde{\mathbf f}^\gamma_{\rm inhom}(\mu_F^2)  \,. 
    \label{eq:hominhom}
\end{align}
The homogeneous term is given by the standard (hadronic) solution of the DGLAP equations
\begin{equation}
    \tilde{\mathbf f}^\gamma_{\rm hom}(\mu_F^2) = \tilde {\mathbf E}(\mu_F^2 \leftarrow \mu_0^2) \tilde{\mathbf f}^\gamma(\mu_0^2) \,,
    \label{eq:solution_hom}
\end{equation}
where $\tilde{\mathbf E}$ is the evolution operator in the EKO framework~\cite{Candido:2022tld}, which solves the homogeneous part of \cref{eq:DGLAP} and which is formally defined in \cref{eq:EKOdef}, and $\tilde{\mathbf f}^\gamma(\mu_0^2)$ is a given boundary condition. 
For the inhomogeneous term, a particular solution with a vanishing initial condition can be written in the following form,
\begin{equation}
    \tilde{\mathbf f}^\gamma_{\rm inhom}(\mu_F^2) = -\tilde{\mathbf E}(\mu_F^2 \leftarrow \mu_0^2) \int\limits_{\mu_0^2}^{\mu_F^2}\! \frac{\diff \mu^2}{\mu^2} \tilde{\mathbf E}(\mu_0^2 \leftarrow \mu^2) \tilde{\mathbf k}(a_s(\mu^2)) \,.
\end{equation}
Using the transitivity of the (hadronic) EKO $\tilde{\mathbf E}$, it can be rewritten as
\begin{equation}
    \tilde{\mathbf f}^\gamma_{\rm inhom}(\mu_F^2) = -\int\limits_{\mu_0^2}^{\mu_F^2}\! \frac{\diff \mu^2}{\mu^2} \tilde{\mathbf E}(\mu_F^2 \leftarrow \mu^2) \tilde{\mathbf k}(a_s(\mu^2)) \,.
    \label{eq:inhom}
\end{equation}
Therefore, we can 
express
the general solution of \cref{eq:DGLAP} in terms of a new  photon evolution operator $\tilde{\mathbf E}^\gamma$
in the following functional form,
\begin{equation}
    \tilde{\mathbf f}^\gamma(\mu_F^2) = \tilde{\mathbf E}^\gamma(\mu_F^2 \leftarrow \mu_0^2)\left[\tilde{\mathbf f}^\gamma(\mu_0^2)\right]
     = \tilde {\mathbf E}(\mu_F^2 \leftarrow \mu_0^2) \tilde{\mathbf f}^\gamma(\mu_0^2) -\int\limits_{\mu_0^2}^{\mu_F^2}\! \frac{\diff \mu^2}{\mu^2} \tilde{\mathbf E}(\mu_F^2 \leftarrow \mu^2) \tilde{\mathbf k}(a_s(\mu^2)) \, .
     \label{eq:gEKOop}
\end{equation}
Note that $\tilde{\mathbf E}^\gamma$ is an affine operator, generalizing the hadronic evolution operator $\tilde{\mathbf E}$ to include the inhomogenous term of the evolution equations. A specific implementation of \cref{eq:gEKOop} was given earlier by ref.~\cite{Gluck:1991ee} and is generalized here. The transitivity of the new photon evolution operator $\tilde{\mathbf E}^\gamma$ follows from the  transitivity of the hadronic EKO $\tilde{\mathbf E}$ operator,
\begin{equation}
    \tilde{\mathbf f}^\gamma(\mu_2^2) = \tilde{\mathbf E}^\gamma(\mu_2^2 \leftarrow \mu_1^2)\left[\tilde{\mathbf E}^\gamma(\mu_1^2 \leftarrow \mu_0^2)\left[\tilde{\mathbf f}^\gamma(\mu_0^2)\right]\right] =  \tilde{\mathbf E}^\gamma(\mu_2^2 \leftarrow \mu_0^2)[\tilde{\mathbf f}^\gamma(\mu_0^2)] \,,
    \label{eq:E_gamma_transit}
\end{equation}
which implies that the separation of the photon PDF into a homogeneous and an inhomogeneous part, see \cref{eq:hominhom}, is merely a mathematically elegant way to solve \cref{eq:DGLAP} with no physical significance.

The discussion so far has implied that all the involved quantities, e.g., the perturbative anomalous dimensions $\gamma(\alpha_s(\mu_F^2))$ and $\mathbf k(\alpha_s(\mu_F^2)$, the strong coupling constant $\alpha_s(\mu_F^2)$, and the photon PDFs $\mathbf f^\gamma(\mu_F^2)$, are  evaluated at a fixed number of light quark flavors $n_f=n_u+n_d$; this corresponds to the so-called
fixed flavor number scheme (FFNS).
When crossing a flavor threshold at $\mu_h$, which is usually set to be the mass of a heavy quark, $\mu_h=m_h$, one transitions from the $n_f$ to the $n_f+1$ scheme,
which requires matching the involved quantities.
Note that 
it 
can be done at any scale, which implies that the number of light flavors $n_f$ and the scale $\mu_F$ are two independent variables~\cite{Barontini:2024xgu}.

Denoting the explicit $n_f$ dependence of the photon PDFs $\mathbf f^\gamma(\mu_F^2)$ by the superscript $(n_f)$, one finds the following general matching condition at $\mu_h$,
\begin{equation}
    \tilde{\mathbf f}^{\gamma,(n_f+1)}(\mu_h^2) = \tilde{\mathbf A}^{(n_f)}(\mu_h^2) \tilde{\mathbf f}^{\gamma,(n_f)}(\mu_h^2) + \tilde{\mathbf A}^{\gamma,(n_f)}(\mu_h^2) 
    \label{eq:matching} \,,
\end{equation}
where $\tilde{\mathbf A}$ refers to the hadronic operator matrix elements~\cite{Buza:1996wv} and $\tilde{\mathbf A}^\gamma$ to their photonic counterparts.
This expression can be obtained by equating the structure function $F_2^{\gamma}$
calculated using \cref{eq:F2_MSbar_3} in the $n_f$ and $n_f+1$ schemes. 
Unlike $F_2^{\gamma}$, which is a physical quantity and thus must be continuous across the heavy flavor threshold, coefficient functions and photon PDFs require a matching procedure encoded in \cref{eq:matching}.

Finally, we present our master formula for the scale evolution of photon PDFs from the initial scale $\mu_0$ with $n_f$ (massless) flavors up to a scale $\mu_1$ with $n_f+1$ flavors, including  PDF matching at $\mu_h$:
\begin{align}
    \tilde{\mathbf f}^{\gamma,(n_f+1)}(\mu_1^2) &=
      \tilde {\mathbf E}^{(n_f+1)}(\mu_1^2 \leftarrow \mu_h^2) \tilde{\mathbf A}^{(n_f)}(\mu_h^2) \tilde {\mathbf E}^{(n_f)}(\mu_h^2 \leftarrow \mu_0^2) \tilde{\mathbf f}^{\gamma,(n_f)}(\mu_0^2) \nonumber\\
     &\hspace{10pt} - \tilde {\mathbf E}^{(n_f+1)}(\mu_1^2 \leftarrow \mu_h^2)\tilde{\mathbf A}^{(n_f)}(\mu_h^2) \int\limits_{\mu_0^2}^{\mu_h^2}\! \frac{\diff \mu^2}{\mu^2} \tilde{\mathbf E}^{(n_f)}(\mu_h^2 \leftarrow \mu^2) \tilde{\mathbf k}^{(n_f)}(a_s^{(n_f)}(\mu^2)) \nonumber\\
     &\hspace{10pt} + \tilde {\mathbf E}^{(n_f+1)}(\mu_1^2 \leftarrow \mu_h^2) \tilde{\mathbf A}^{\gamma,(n_f)}(\mu_h^2) \nonumber \\
     &\hspace{10pt} -\int\limits_{\mu_h^2}^{\mu_1^2}\! \frac{\diff \mu^2}{\mu^2} \tilde{\mathbf E}^{(n_f+1)}(\mu_1^2 \leftarrow \mu^2) \tilde{\mathbf k}^{(n_f+1)}(a_s^{(n_f+1)}(\mu^2)) \,.
     \label{eq:evolution_master}
\end{align}
While the first line is directly analogous to the scale evolution of proton PDFs, the remaining terms 
are specific to photon PDFs and
originate from the point-like contribution due to the $\gamma \to q {\bar q}$ coupling.
Note that \cref{eq:evolution_master} generalizes \cref{eq:E_gamma_transit} by adopting the variable flavor number scheme (VFNS) instead of the FFNS one.

In our analysis, we restrict ourselves to NLO accuracy, where the matching conditions, \cref{eq:matching}, simplify\footnote{This is no longer true, when one considers intrinsic heavy quark contributions~\cite{Ball:2022qks}.} and yield
\begin{equation}
\tilde{\vb A}^{(n_f)}(m_h^2) = \vb 1 \qand \tilde{\vb A}^{\gamma,(n_f)}(m_h^2) = 0 \,,
\label{eq:matchin2}
\end{equation}
which leads to corresponding simplifications in  \cref{eq:evolution_master}.

The numerical implementation of \cref{eq:evolution_master} is realized through the existing EKO framework for the purely hadronic part (first line) and a new open-source program $\gamma$\texttt{EKO}~\cite{felix_hekhorn_2025_16032673} (pronounced [geko]) for the general point-like solution of photon PDF evolution (the last three lines).
Equation~(\ref{eq:evolution_master}) and the new $\gamma$\texttt{EKO} code developed for its solution are among the original results of our work.
Further details on our implementation of scale evolution of photon PDFs can be found in \cref{app:geko}, where we also present the explicit solution at LO accuracy and discuss solution strategies at NLO accuracy and beyond.

%% file: sec_meth.tex
The goal of this work is to determine photon PDFs $f_{j}^{\gamma}$ by applying the machinery of global QCD fits
to the available data on the $F_2^{\gamma}$ structure functions.
This is based on three main components: (i) the LO and NLO QCD analysis of $F_2^{\gamma}$
in terms of $f_{j}^{\gamma}$, discussed in section~\ref{sec:theory},
(ii) statistical methods allowing one to quantify the quality of extraction of  $f_{j}^{\gamma}$ from $F_2^{\gamma}$, which we discuss in this section, and (iii) comprehensive experimental data on
$F_2^{\gamma}$, which we summarize later in \cref{sec:data}. 

\subsection{Overview of Monte Carlo framework}
\label{sec:MC}

In our analysis, as a statistical tool, we use the framework of Monte Carlo (MC) replicas for photon PDFs. The average over all replicas gives the 
central PDFs, while replica deviations from the average serve as an estimate of PDF uncertainties.
For a detailed mathematical review of the MC replica method, we refer the reader to refs.~\cite{Costantini:2024wby,DelDebbio:2021whr} and limit ourselves here to a short summary.

Denoting collectively the experimentally measured data on the photon structure function $F_2^{\gamma}$ as $\{D_j\}_{j=1..N_\text{dat}}$, where $N_\text{dat}$ is the number of data points, 
one 
commonly
assumes that they have Gaussian uncertainties $\{\sigma_j\}_{j=1..N_\text{dat}}$.
Because of the linear mapping between $F_2^{\gamma}$ and the photon PDFs 
in the form of FK tables, see \cref{eq:factFK}, it implies that $\vb f^{\gamma}(x,Q^2)$ at a given $x$
also obey a Gaussian distribution.
We further assume that the photon PDFs at an initial scale $Q_0$
(we take $Q_0^2 = 1$ GeV$^2$ in our analysis) can be parameterized using a set of fitting parameters $\vec \theta$, 
\begin{equation}
    \vb f^\gamma(x,Q_0^2) = \vb f_0^\gamma(x; \vec \theta)\,.
\end{equation}
Thus, using the FK tables, eq.~(\ref{eq:factFK}), a theory prediction $T_j$ corresponding to a data point $D_j$ can be written as
\begin{equation}
    T_j(\vec \theta) = \vb{FK}_j \otimes \vb f^\gamma_0(\vec \theta) + \mathrm{FK}_j^\gamma \,,
\end{equation}
where $\vb{FK}_j$  and $\mathrm{FK}_j^\gamma$ correspond to the necessary hadronic  and photonic FK tables, respectively.

In our analysis, we generate a set of $N_\text{rep}=100$ data replicas, where each replica $k$ corresponds to a specific list of pseudo-data $\{D_{j,(k)}\}_{j=1..N_\text{dat}}^{k=1..N_\text{rep}}$.
The pseudo-data points are drawn from a Gaussian distribution, which is centered at the actual measured data points $D_j$, with their standard deviation being equal to the uncertainty $\sigma_j$. Replicas with negative $F_2^\gamma$ are redrawn to ensure the physicality of the replica.

For each pseudo-data replica, we then find a corresponding PDF replica by minimizing the loss function, $L(\vec\theta_{(k)})$ with respect to the PDF parameters $\vec\theta_{(k)}$.
Specifically, for each replica $k$ and data point $j$, we introduce the residuals
\begin{equation}
    r_{j,(k)}(\vec\theta_{(k)}) = D_{j,(k)}-T_j(\vec\theta_{(k)}) \,,
\end{equation}
which measure the distance between the pseudo-data and our theory predictions. 
We then define the loss function
\begin{equation}
    L_{(k)}(\vec\theta_{(k)}) = \sum_{j=1}^{N_\text{dat}} \rho\Bigg(\qty(\frac{r_{j,(k)}(\vec \theta_{(k)})}{\sigma_j})^2\Bigg) \,,
    \label{eq:loss}
\end{equation}
where $\rho$ is 
the
so-called cost function.
For the linear case $\rho(z) = z$, the loss function reduces to the familiar $\chi^{2}_{(k)}$ equation,
\begin{equation}
    \chi^{2}_{(k)} (\vec\theta_{(k)}) = \sum_{j=1}^{N_\text{dat}} \Bigg(\frac{D_{j,(k)}-T_j(\vec\theta_{(k)})}{\sigma_{j}}\Bigg)^2\,.
    \label{eq:repchi2}
\end{equation}
Further, to reduce the impact from outliers, we 
replace 
the linear cost function
by
the so-called \texttt{soft\_l1} approximation~\cite{Virtanen:2019joe,dembinski_2025_17448283},
\begin{equation}
    \rho(z) = 2 \Big(\sqrt{1+z} -1\Big)\,.
\end{equation}
This cost function approximates the regular linear cost function for small deviations, $r_{j,(k)}(\vec\theta_{(k)})\ll \sigma_j$, but suppresses the contributions from larger deviations, $r_{j,(k)}(\vec\theta_{(k)}) > \sigma_j$. 

To minimize the loss function, \cref{eq:loss}, in our analysis, we tested two minimization algorithms: \texttt{least\_squares} from SciPy~\cite{Virtanen:2019joe} and its counterpart from \texttt{iminuit}~\cite{dembinski_2025_17448283}. Both minimization algorithms converge towards similar minima across all replicas, but the \texttt{iminuit} package yields a slightly better $\chi^2_{(k)}$ and converges significantly faster. Thus we use the latter for this study.
Moreover, we follow the NNPDF approach and discard replicas, whose $\chi^{2}_{(k)}$ diverge by more than $4\sigma$ from the ensemble average~\cite{NNPDF:2021njg}.
Eventually, we obtain a set of optimal fitting parameters $\vec \theta^\text{opt}_{(k)}$, each corresponding to a given set of pseudo-data points, which form our ensemble of PDF replicas.
The best-fit, central PDFs are then obtained by averaging over all the replicas,
\begin{equation}
    \vb f^\gamma_{0,c}(x) = \frac 1 {N_\text{rep}} \sum_{k=1}^{N_\text{rep}} \vb f_0^\gamma\qty(x;\vec \theta_{(k)}^\text{opt}) \,.
    \label{eq:central}
\end{equation}

A few important comments regarding our procedure for determining PDFs 
are in order.
First, the only significant object is the central PDF $\vb f^\gamma_{0,c}$, 
while a single replica does not represent any 
statistically
relevant information. 
In other words, it is the entire ensemble
that
contains all important information.
Second, since the averaging over the replicas in \cref{eq:central} is performed at a given momentum fraction $x$, it means that the central PDF $\vb f^\gamma_{0,c}$ may or may not be given by the same functional form as the replicas.
In particular, if there is a non-linear dependence on the fitting parameters, which is in practice almost always the case and which is also present in our case, the central PDF has no simpler representation than that given by \cref{eq:central}.
Moreover, the non-linearity also implies that the fitting parameters $\vec\theta_{(k)}$ do not necessarily obey a Gaussian distribution, since we only assume that the PDF itself at a given momentum fraction $x$ is a Gaussian distribution.

To calculate uncertainties of our extracted PDFs, we used two standard statistical estimators: 
the standard deviation and the \SI{68}{\percent} confidence interval (CI).
Specifically, for a set of values of photon PDFs at given $x$, $Q^2$, parton flavor, and replica $k$ and for a set of values of $F_2^{\gamma}$ at given $x$, $Q^2$, and replica $k$, which we collectively denote $\{g_{(k)}\}^{k=1..N_\text{rep}}$, the standard deviation is defined by averaging over the replicas,
\begin{equation}
    \Delta g_\sigma^2 = \frac 1 {N_\text{rep}} \sum_{k=1}^{N_\text{rep}}\left[ \left(\frac 1 {N_\text{rep}}\sum_{k'=1}^{N_\text{rep}}g_{(k')}\right) - g_{(k)} \right]^2 \,.
    \label{eq:st_div}
\end{equation}
In addition, the \SI{68}{\percent} confidence interval $[\Delta g_-,\Delta g_+]$ is defined by sorting 
through
the replicas 
and  keeping those satisfying the following relation,
\begin{equation}
    \frac{1 - 0.68}{2} = \frac 1 {N_\text{rep}} \sum_{k=1}^{N_\text{rep}}\Theta(\Delta g_- - g_{(k)}) = \frac 1 {N_\text{rep}} \sum_{k=1}^{N_\text{rep}}\Theta(g_{(k)} - \Delta g_+) \,,
    \label{eq:CI}
\end{equation}
where $\Theta$ is the usual Heaviside 
step
function.
Applying these two statistical estimators to our photon PDFs at fixed $x$, $Q^2$, and flavor
and to our theoretical calculations for $F_2^{\gamma}$ at given $x$ and $Q^2$,
we obtain the corresponding photon PDF and $F_2^{\gamma}$ uncertainty bands (confidence intervals).
We have checked explicitly that the number of replicas, which we use ($N_{\rm rep}=100$), is sufficient to obtain robust error estimates.

Ultimately, we can define our figure-of-merit, the central $\chi^2$ using the following relation,
\begin{equation}
    \chi^2 = \sum_{j=1}^{N_\text{dat}} \Bigg(\frac{D_{j}-\vb{FK}_j\otimes \vb f^\gamma_{0,c} - \text{FK}^\gamma_j}{\sigma_{j}}\Bigg)^2 \,.
    \label{eq:chi2}
\end{equation}
Note that while $\chi^{2}_{(k)} (\vec\theta_{(k)})$  in \cref{eq:repchi2} and $\chi^2$ in \cref{eq:chi2} resemble each other, there is no direct relation between %
them.
While the former is used for the minimization of each replica, the latter involves averaging over the entire ensemble of replicas.
Thus, it is the central $\chi^2$ of \cref{eq:chi2} that carries a physical interpretation and is used in our analysis to access the quality of our fits and compatibility of different data sets. 

\subsection{Initial conditions for photon PDFs}

In any global QCD analysis of PDFs
a choice of the functional form of initial conditions and the input scale $Q_0$ introduce a certain bias and model dependence. 
We parametrize the initial conditions for photon PDFs 
$f_0^\gamma(x) \equiv f^\gamma(x,Q^2=Q_0^2)$
at the input scale $Q_0 = \SI{1}{\GeV}$ using the following form,
\begin{align}
\frac{1}{\alphaem} xu_0^\gamma(x) = \frac{1}{\alphaem} x\bar u_0^\gamma(x) &= N_u x^{a_u}(1-x)^{b_u}  \,,  \nonumber\\  
d_0^\gamma(x)  =   \bar d_0^\gamma(x) &=  u_0^\gamma(x) \,, \nonumber\\ 
    s_0^\gamma(x) = \bar s_0^\gamma(x) &=K_s \, u_0^\gamma(x) \,,  \nonumber\\
  \frac{1}{\alphaem}  xg_0^\gamma(x) &= N_gx^{a_g}(1-x)^{b_g} \,,
    \label{eq:input_param}
\end{align}
where
we impose exact isospin symmetry and zero valence distributions $q_j^\gamma - \bar q_j^\gamma = 0$. In addition we assume a vanishing charm quark distribution at the initial scale, $c_0^\gamma(x) = \bar c_0^\gamma(x)=0$, which is then generated purely perturbatively for $Q^2 > m_c^2$ through evolution.
The parametrization of \cref{eq:input_param} implies a hadron-like form of the input photon PDFs, which we consider to be a reasonable assumption at low scales, when the photon point-like component can be neglected.
Note that scale evolution tends to gradually break down
the isospin symmetry assumed at the input scale, leading to $u^{\gamma}(x,Q^2) > d^{\gamma}(x,Q^2)$ for $Q^2 > Q_0^2$, which becomes especially prominent for very large values of $x> 0.8$.

The form used in \cref{eq:input_param} is inspired by the photon-vector meson
connection of the VMD model, which can be realized througth the following frequently used relation
\begin{equation}
\frac{1}{\alphaem} f_{j,0}^\gamma(x) \sim \sum_{V=\rho, \omega, \phi}\frac{1}{f_V^2}  f_j^V(x) \sim  f_j^{\pi}(x) \,,
\label{eq:input_vmd}
\end{equation}
where $f_j^V(x)$ and $f_j^{\pi}(x)$ are the vector meson and pion PDFs, respectively,
and $f_V$ are the photon-vector meson coupling constants.

Note that it is customary and convenient to present photon PDFs divided by the fine-structure constant $\alpha_{\rm em}$ because experimental data on the photon structure function is commonly reported for
$F_2^{\gamma}/\alpha_{\rm em}$. 
Further, to the leading order in $\alpha_{\rm em}$, this allows one to factor out 
$\alpha_{\rm em}$ from both the expression for $F_2^{\gamma}$, \cref{eq:F2_MSbar_3}, and the scale evolution equations for photon PDFs, \cref{eq:ev_matrix}.

In our analysis, we treat the quark parameters $(N_u, a_u, b_u)$ and the gluon parameters
$(N_g,a_g)$ as 5 free parameters of the fit, $\vec\theta =\{N_u, a_u, b_u, N_g, a_g\}$.
The two remaining parameters are taken to be constant, $b_g=3$ and $K_s=0.3$~\cite{Slominski:2005bw}, 
with their values motivated by 
VMD-inspired relation of Eq.~(\ref{eq:input_vmd}).
First, 
relating
the gluon distribution in the real photon 
to
that in the pion and using the quark counting rules for pion PDFs in the $x \to 1$ limit~\cite{Blankenbecler:1974tm,Farrar:1975yb}, one finds $b_g=3$.
Second, 
substituting the values for the coupling constants $f_{\rho}$, $f_{\omega}$, and $f_{\phi}$~\cite{Klein:1999qj} into Eq.~(\ref{eq:input_vmd}), 
one estimates that strange quarks contribute approximately 15\% to the photon hadronic structure.
Thus, the strange quark distribution in the photon is expected to be suppressed compared to the up and down quarks by the factor of $K_s\approx0.3$.
We have checked that varying 
$b_g$ and $K_s$ within the
$2 \leq b_g \leq 4$ and  $0.2 \leq K_s \leq 0.4$ ranges does not result in appreciable changes in the quality of our fits.

We also investigated other, more flexible parameterization forms, for example, using a mixture of the hadron-like and point-like contributions~\cite{Slominski:2005bw}, or leaving all parameters in \cref{eq:input_param} free.
These did not result in significant fit improvements and mostly led to more unstable and outlying replicas, leading to distinctly different convergent solutions of the fit parameters.
In summary, while our ansatz in \cref{eq:input_param} for
the up and down quark distributions is sufficiently flexible, the uncertainties on the
strange quark and gluon distributions are bound to be underestimated. This %
reflects the limited sensitivity of the fitted photon structure function $F_2^{\gamma}$ to the quark flavor separation and to the gluon distribution at large values of $x$.

The values of $\alpha_s$, $\alpha_{\rm em}$, and the heavy heavy quark masses, which serve as standard theoretical parameters of our fit, are listed in \cref{tab:th_values}.
Note that we use the two-loop (one-loop) expression for $\alpha_s$ along with the 
$\alpha_s(M_Z) = 0.118$ normalization at NLO (LO) accuracy.
\begin{table}[H]
    \caption{The values of $\alpha_s$,  $\alphaem$, and the heavy heavy quark masses used in our fit.}
    \centering
    \begin{tabular}{c c}\toprule
         $\alpha_s(M_Z)$&0.118  \\
         $1/\alpha_{\text{em}}$ & 137\\
         $m_c$ & \SI{1.30}{\GeV} \\
         $m_b$ & \SI{4.75}{\GeV} \\
         $m_t$ & \SI{172}{\GeV} \\
         \bottomrule
    \end{tabular}
    \label{tab:th_values}
\end{table}

In principle, one could further constrain the free parameters in \cref{eq:input_param}, in particular the gluon normalization parameter $N_g$, using the proposed momentum sum rule (MSR) for photon PDFs~\cite{Schuler:1995fk,Frankfurt:1996nz,Frankfurt:1995bd},
\begin{equation}
\frac{1}{\alpha_{\rm em}} \int_0^1 dx x \left(\Sigma^{\gamma}(x,Q^2)+g^{\gamma}(x,Q^2)\right)=c+ \sum_{j}^{n_f} e_{q_j}^2 \frac{1}{\pi} \ln \left(\frac{Q^2}{\tilde{Q}_0^2}\right) \,.
\label{eq:sr}
\end{equation}
The first and second terms on the right correspond to the hadronic and point-like 
contributions, respectively:
the parameter $c$ can be estimated by using the VMD model and
the parameter $\tilde{Q}_0$
plays the role of an effective scale separating the VMD and point-like (anomalous) contributions. 
However, the uncertainties in these parameters as well as the question on the scheme dependence make it impractical 
to use the MSR, \cref{eq:sr}, for constraining the parametrization of \cref{eq:input_param} in our analysis. 

%% file: sec_data.tex
\begin{table}[ht!]
    \centering
    \caption{Photon structure function $F_2^{\gamma}$ data from electron-positron colliders used in the present analysis. Data sets marked with an asterisk (*) are not included in the fit.}
    \begin{tabular}{c l r r r r}\toprule
        Collider &  Data set & Reference & $Q^2$ [GeV$^2$]& \ $N_\text{dat}$ & HEPData\\
        \midrule
        \multirow{10}{*}{{\rotatebox[origin=c]{90}{LEP-1, LEP-2}}}& ALEPH, 1999 & \cite{ALEPH:1999vwa} & [9.9, 284.0] & 11 &\cite{hepdata.49109.v1/t8}\\
        & ALEPH, 2003 & \cite{ALEPH:2003pxe} & [17.3, 67.2] & 16 & \cite{hepdata.43218.v1/t1} \\
        & DELPHI, 1996 & \cite{DELPHI:1995fid} & 12.0 & 4 & \cite{hepdata.47867.v1/t1}\\
        & L3, 1998 & \cite{L3:1998bfn} & [1.9, 5.0] & 12 & \cite{hepdata.49392.v1/t1}\\
        & L3, 1999 & \cite{L3:1998ryp}
        & [10.8 23.1] & 11 & \cite{hepdata.49323.v1/t1}\\
        & L3, 2000~* & \cite{L3:2000leb} & 120.0 & 5 & \cite{hepdata.49964}\\
        & OPAL, 1994 & \cite{OPAL:1993srf} & [5.9, 14.7] & 7 & \cite{hepdata.48474.v1/t1}\\
        & OPAL, 1996 & \cite{OPAL:1996tah} & [7.5, 135.0] & 10 & \cite{hepdata.47770.v1/t2}\\
        & OPAL, 1998 & \cite{OPAL:1997day} & [9.0, 59.0] & 14 & \cite{hepdata.47450.v1/t2}\\
        & OPAL, 1998 & \cite{OPAL:1997wop} & [1.86, 3.76] & 8 & \cite{hepdata.49560.v1/t2}\\
        & OPAL, 2000 & \cite{OPAL:2000nfx} & [1.9, 17.8] & 22 & \cite{hepdata.49907}\\
        & OPAL, 2002~* & \cite{OPAL:2002vci} & [12.1, 780.0] & 13 & \cite{hepdata.49744.v1/t3}\\
        \midrule
        \multirow{4}{*}{{\rotatebox[origin=c]{90}{PETRA}}}& JADE, 1984 & \cite{JADE:1984fxa} & [24.0, 100.0] & 8 & -\\
        & PLUTO, 1984 & \cite{PLUTO:1984pmb} & [2.4, 9.2] & 9 & \cite{hepdata.30545.v1/t3}\\
        & PLUTO, 1986 & \cite{PLUTO:1986dcs} & 
        45.0 & 4 & \cite{hepdata.33588.v1/t2}\\
        & TASSO, 1986 & \cite{TASSO:1986ats} & 23.0 & 5 & \cite{hepdata.15858.v1/t1}\\
        \midrule
        \multirow{5}{*}{{\rotatebox[origin=c]{90}{TRISTAN}}}&&&&\\
        & AMY, 1995 & \cite{AMY:1995idu} & [73.0, 390.0] & 5 & \cite{hepdata.38361.v1/t2}\\
        & AMY, 1997 & \cite{AMY:1997hbe} & 6.8 & 3 & \cite{hepdata.28320.v1/t1}\\
        & TOPAZ, 1994 & \cite{TOPAZ:1994xqa} & [5.1, 80.0] & 8 & \cite{hepdata.38377.v1/t3}\\
        &&&&\\
        \midrule
        \multirow{3}{*}{{\rotatebox[origin=c]{90}{SLAC}}}&&&&\\
        & TPC/Two-Gamma, 1986~* & \cite{TPCTwoGamma:1986ycp} & [0.24, 5.1] & 22 & \cite{hepdata.15803}\\
        &&&&\\
        \midrule        
        \multicolumn{4}{r}{\text{Total points used}}& 157 & \\
        \bottomrule
    \end{tabular}
    \label{tab:data}
\end{table}

\begin{figure}[t]
    \centering
    \includegraphics[width=0.9\linewidth]{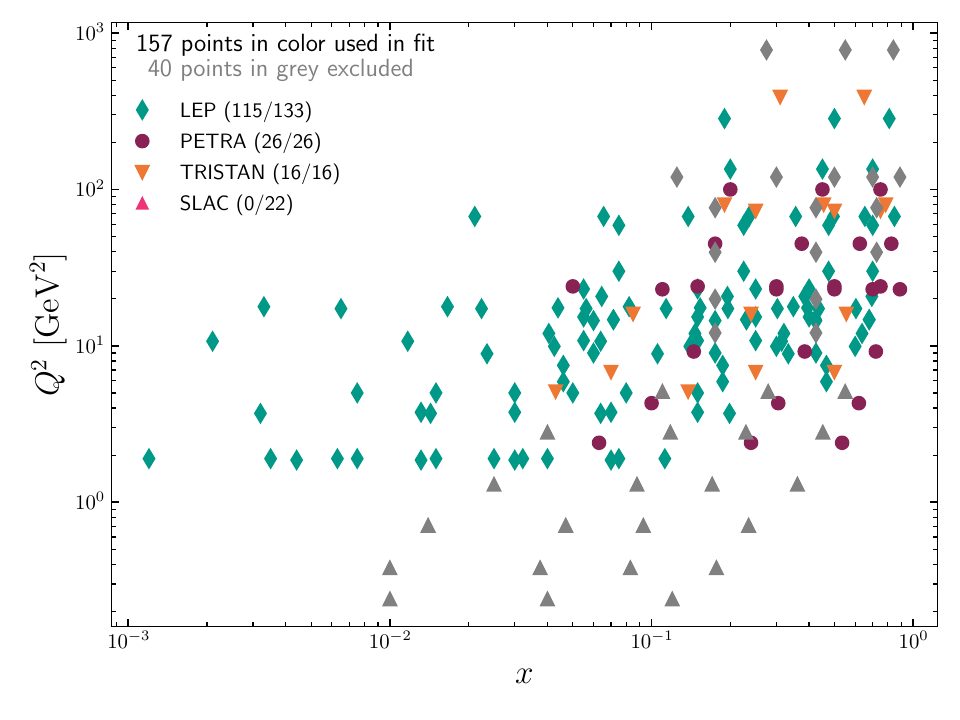}
    \caption{The kinematic coverage in the $x\text{-}Q^2$ plane of the world data on the photon structure function
    $F_2^{\gamma}$ from $e^+e^-$ collisions. Colored symbols  (diamonds, circles, inverted triangles) show the data points used in our global fit, while the data in grey (diamonds, triangles) are excluded from our analysis. The data are labeled by the collider name, indicating also the number of used and total available points
    in the parenthesis.
    }
    \label{fig:kin}
\end{figure}

The available data on the photon structure function $F_2^{\gamma}$ come from four 
electron-positron colliders: 
the largest amount 
is
from the Large Electron Positron (LEP) collider at CERN, while a smaller fraction comes from the Positron–Electron Tandem Ring Accelerator (PETRA) at DESY, the Transposable Ring Intersecting Storage Accelerator in Nippon (TRISTAN) collider at KEK, and the Stanford Linear Accelerator Center (SLAC).
While a comprehensive source of available data can be found in ref.~\cite{Nisius:1999cv}, whenever possible, we rely on the original publications from the HEPData database~\cite{Maguire:2017ypu}, also to avoid possible transcription errors.
We give an overview of the available data in \cref{tab:data} and the respective kinematic ranges in the $x\text{-}Q^2$ plane in \cref{fig:kin}.

In the following, we give further details on our treatment of specific data sets.
\begin{itemize}
\item Most of the data from the TPC/Two-Gamma collaboration at SLAC~\cite{TPCTwoGamma:1986ycp} lies below $Q^2 < \SI{1}{\GeV^2}$ and, thus, is not used in our analysis.

\item When data sets have asymmetric systematic uncertainties, specifically, the OPAL measurement of refs.~\cite{OPAL:1996tah,OPAL:1997day,OPAL:1997wop,OPAL:2002vci}, we symmetrize them using the D'Agostini prescription~\cite{DAgostini:2004kis}.

\item L3~\cite{L3:1998bfn, L3:1998ryp}. 
Measurements are presented in two sets: set 1 unfolded with the PHOJET Monte Carlo and set 2 unfolded with TWOGAM.
We utilize results and the corresponding statistical errors from set 1, while the systematic error is the quadratic sum of the systematic error obtained for set 1 and the difference between the results of set 1 and 2.

\item PLUTO~\cite{PLUTO:1984pmb}.
The systematic error is calculated at 10\% relative to the value of the photon structure function for points $0.3 < x < 0.8 $, 15\% for $0.2 < x < 0.3$, and 25\% for $x < 0.2$.
We use the fully inclusive measurement.
      
\item PLUTO~\cite{PLUTO:1986dcs}. The systematic error is flat 10\%.
We use the fully inclusive measurement.
      
\item TASSO~\cite{TASSO:1986ats}.
Systematic uncertainties are taken from ref.~\cite{Nisius:1999cv} and correspond to around 19\% relative uncertainty.

\item
We find that the L3 2000~\cite{L3:2000leb}, 
the
OPAL 2002~\cite{OPAL:2002vci}, %
and the SLAC data points~\cite{TPCTwoGamma:1986ycp} 
which pass the $Q^2 \geq 1$ GeV$^2$ cutoff, yield 
unjustifiably
large contributions to our figure-of-merit $\chi^2$, \cref{eq:chi2}.
Therefore, we have excluded them from our fit. 
\end{itemize}

In summary, in our analysis, we use 157 data points (115 LEP, 26 PETRA, and 16 TRISTAN points) 
spanning the following range in the $x\text{-}Q^2$ 
plane: $10^{-3} \leq  x \leq 0.98$ and $1.86 \leq Q^2 \leq 390$ GeV$^2$.
In \cref{fig:kin}, they are represented by colored symbols and are labeled by the names of the respective colliders, indicating also the number of used and total points in the parenthesis. 
For completeness, we also show the excluded 20 LEP and 22 SLAC data points by gray symbols.

%% file: sec_results.tex
In this section
we present details of our numerical analysis, including quality of the fits and comparison to selected $F_2^{\gamma}$ data 
used in our fits, the 
resulting
photon PDFs for different flavors at the initial scale and after the scale evolution, and also comparison with several 
photon PDFs available in the literature.

\subsection{Fit quality}
\label{subsection:FitQuality}
\input{sec_results_fitquality}

\subsection{VALO1.0 photon PDFs}
\label{subsec:ThePDFs}

\input{sec_results_ThePDFs}
\subsection{Comparison to selected photon PDFs available in the literature}
\label{sec:PDFcomparisons}

\input{sec_results_PDFcomparisons}
\subsection{Deliverables}

Our final photon PDF are available from the \texttt{LHAPDF} index under the following names
\begin{itemize}
    \item \texttt{VALO10\_LO} for our LO fit
    \item \texttt{VALO10\_NLO\_DISg} for our NLO fit in the DIS$_\gamma$ scheme
    \item \texttt{VALO10\_NLO\_MSbar} for our NLO fit in the $\overline{\rm MS}$ scheme
\end{itemize}
and 
via zenodo~\cite{chithirasreemadam_2026_19709212}. All PDFs are available as 100 replicas in addition to the central fit result.
Note that although here we show plots which are factorized by the electromagnetic coupling, $\alphaem$, we provide the grids as $xf_j(x,Q^2)$ without the division with the EM coupling to follow the LHAPDF standard.

Our fitting framework \texttt{VALOfitter}, which has been used to produce the results of this study can be obtained from the repository\footnote{\url{https://gitlab.jyu.fi/ilmahele/VALOfitter}} 
and
via zenodo~\cite{hekhorn_2026_19694026}. The photon evolution code, \texttt{gEKO}, is available from the repository\footnote{\url{https://github.com/felixhekhorn/geko}}
and
via zenodo~\cite{felix_hekhorn_2025_16032673}.

%% file: sec_results_fitquality.tex
The goodness
of our global QCD fits to the data on $F_2^{\gamma}$ at LO and NLO accuracy is summarized in \cref{tab:chi2_datasets}.
It presents the values of $\chi^2/N_{\rm dat}$ for individual data sets used in our analysis, i.e., the individual contributions to the central $\chi^2$ divided by the used number of data points $N_{\rm dat}$,
along with the total $\chi^2$ per degree of freedom, $\chi^2/\text{\scriptsize DOF}$.
As usual, the number of degrees of freedom ${\rm DOF}$ is defined as the number of data points minus the number fitted parameters, i.e., ${\rm DOF}=157-5=152$.  

\begin{table}[ht!]
    \centering
    \caption{
    Contributions of individual data sets used in our global QCD fits at LO and NLO accuracy
    to the central $\chi^2$ divided by the number of used data points, $\chi^2/N_{\rm dat}$ (second and third columns),
    and the total $\chi^2$ per degree of freedom, $\chi^2/\text{\scriptsize DOF}$.}
    \begin{tabular}{l c c c }\toprule
        \multirow{1}{*}{Data set}& LO $\chi^2/N_{\rm dat}$ & NLO $\chi^2/N_{\rm dat}$ & $N_\text{dat}$\\
        \midrule
        LEP ALEPH 1999 \cite{ALEPH:1999vwa} & 0.47 & 0.55 & 11 \\
        LEP ALEPH 2003 \cite{ALEPH:2003pxe} & 0.60 & 0.79 & 16\\
        LEP DELPHI 1996 \cite{DELPHI:1995fid} & 0.72 & 0.97 & 4 \\
        LEP L3 1998 \cite{L3:1998bfn} & 0.69 & 0.58 & 12\\
        LEP L3 1999 \cite{L3:1998ryp} & 0.72 & 0.52 & 11\\
        LEP OPAL 1994 \cite{OPAL:1993srf} & 0.97 & 1.44 & 7\\
        LEP OPAL 1996 \cite{OPAL:1996tah} & 0.24 & 0.23 & 10\\
        LEP OPAL 1997 \cite{OPAL:1997day} & 0.23 & 0.20 & 14\\
        LEP OPAL 1997 \cite{OPAL:1997wop} & 0.95 & 1.03 & 8\\
        LEP OPAL 2000 \cite{OPAL:2000nfx} & 1.00 & 1.37 & 22\\
        PETRA JADE 1984 \cite{JADE:1984fxa} & 1.93 & 2.44 & 8\\
        PETRA PLUTO 1984 \cite{PLUTO:1984pmb} & 0.22 & 0.20 & 9 \\
        PETRA PLUTO 1986 \cite{PLUTO:1986dcs} & 0.57 & 0.43 & 4 \\
        PETRA TASSO 1986 \cite{TASSO:1986ats} & 1.24 & 1.58 & 5\\
        TRISTAN AMY 1995 \cite{AMY:1995idu} & 0.84 & 1.26 & 5\\
        TRISTAN AMY 1997 \cite{AMY:1997hbe} & 0.74 & 0.50 & 3\\
        TRISTAN TOPAZ 1994 \cite{TOPAZ:1994xqa} & 1.91 & 1.99 & 8 \\
        \midrule
        $\chi^2/\text{\scriptsize DOF}$ & 0.81 & 0.94 & 157 \\
        \bottomrule
    \end{tabular}
    \label{tab:chi2_datasets}
\end{table}

As one can see from \cref{tab:chi2_datasets}, both LO and NLO fits successfully converge, corresponding to $\chi^2/\text{\scriptsize DOF} = 0.81$ at LO and $\chi^2/\text{\scriptsize DOF} = 0.94$ at NLO, respectively.
This marginal increase indicates that our initial conditions may not be sufficiently flexible to accommodate faster scale evolution of photon PDFs and direct sensitivity of $F_2^{\gamma}$ to the gluon distribution at the NLO accuracy.
The individual data sets also demonstrate the good convergence of our LO and NLO fits since for most of them $\chi^2/N_{\rm dat} \lesssim 1$.
The notable exceptions are JADE 1984~\cite{JADE:1984fxa}, TOPAZ 1994~\cite{TOPAZ:1994xqa} and possibly TASSO 1986~\cite{TASSO:1986ats},  which are incidentally the three oldest data sets,
which we include in our fits.

\begin{figure}[t!]
    \centering
        \includegraphics[scale=0.47]{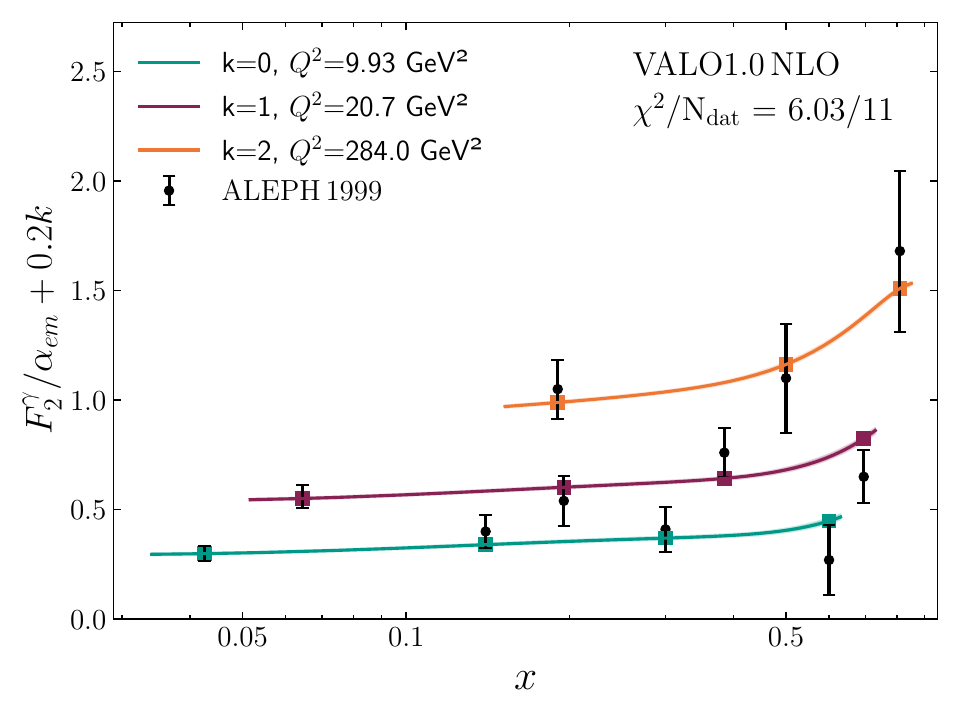}%
        \includegraphics[scale=0.47]{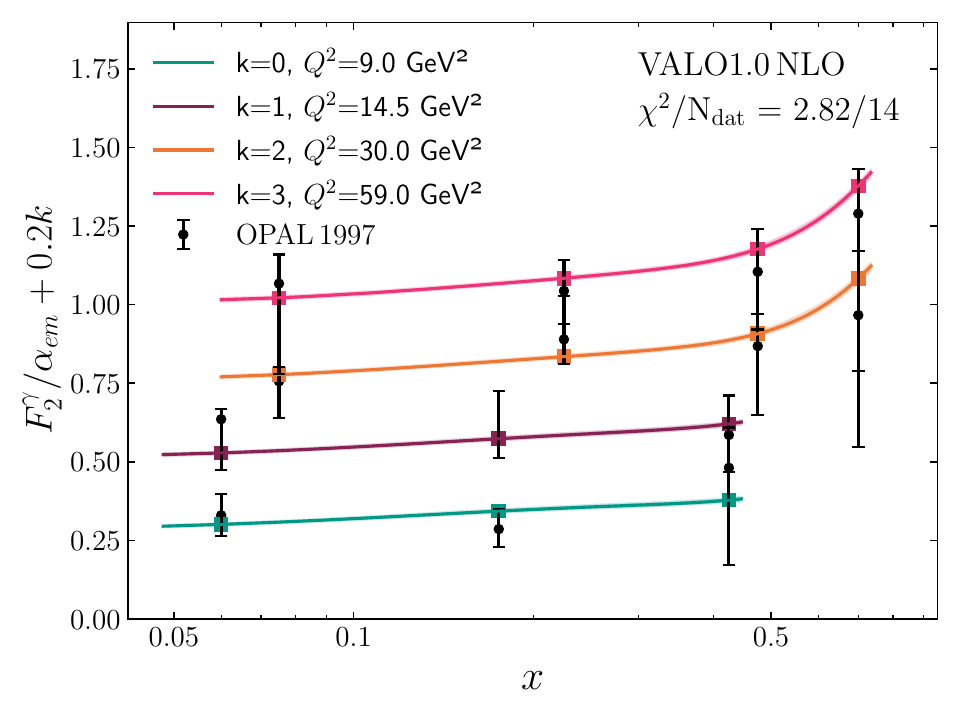}
        \includegraphics[scale=0.47]{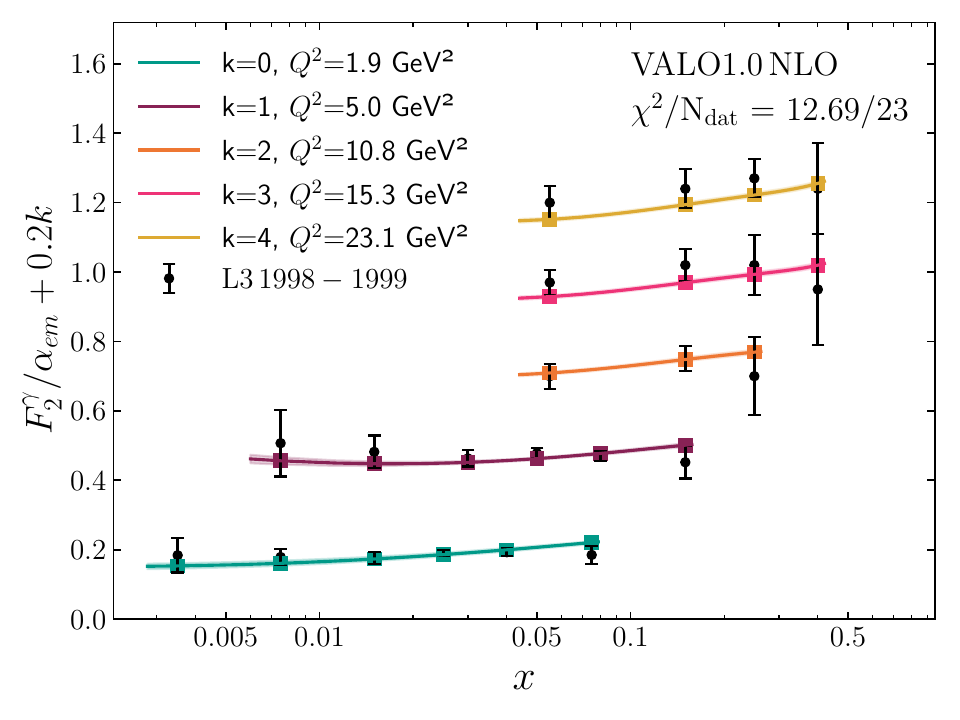}%
        \includegraphics[scale=0.47]{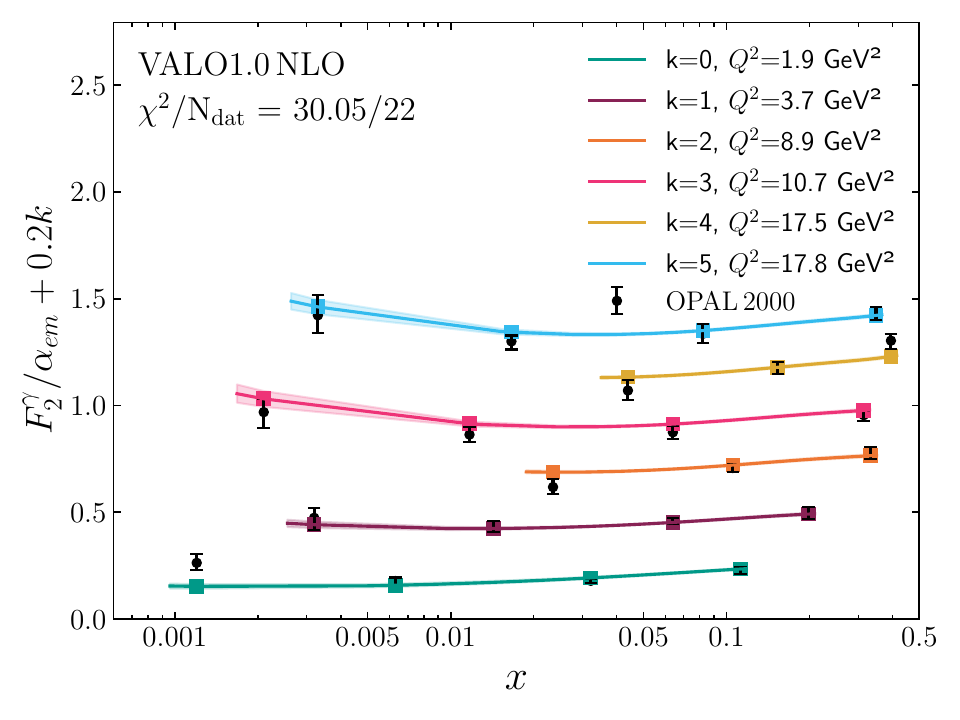}
        \includegraphics[scale=0.47]{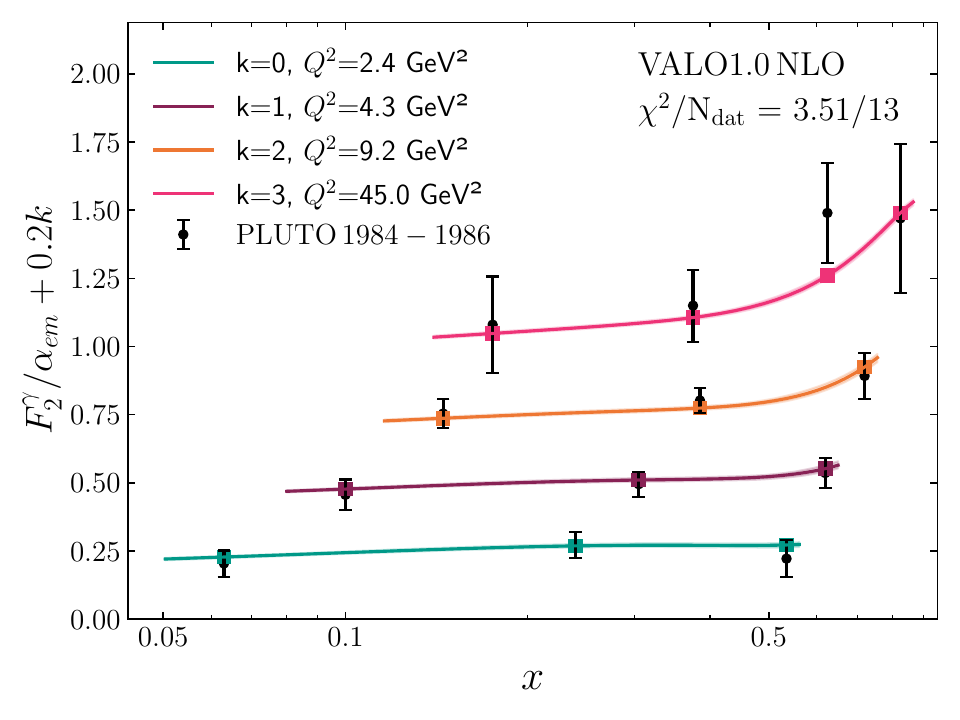}%
        \includegraphics[scale=0.47]{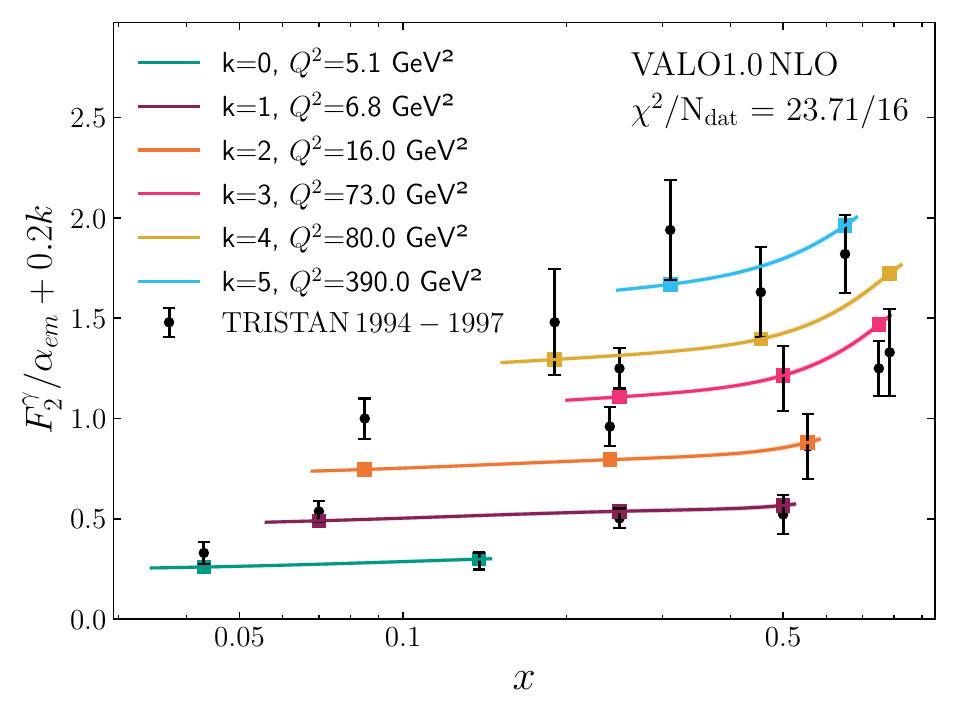}  
    \caption{
    The photon structure function $F_2^{\gamma}$ as a function of $x$ at fixed values of $Q^2$: NLO results using VALO1.0 photon PDFs vs.~selected experimental points used in the fit. The error bars correspond to systematic and statistical uncertainties added in quadrature. The offset factor of 
     $0.2k$, where $0 \leq k \leq 5$, is introduced for legibility.}
    \label{fig:datatheory}
\end{figure}

The good quality of our fits is also illustrated graphically in \cref{fig:datatheory}, 
which compares theory 
results for
the photon structure function 
$F_2^{\gamma}$ as a function of $x$ at fixed values of $Q^2$
with the measured values of $F_2^{\gamma}$ from selected data sets (ALEPH 1999~\cite{ALEPH:1999vwa}, OPAL 1997~\cite{OPAL:1997day,OPAL:1997wop}, L3~\cite{L3:1998bfn,L3:1998ryp,L3:2000leb}, OPAL 2000~\cite{OPAL:2000nfx}, PLUTO~\cite{PLUTO:1984pmb,PLUTO:1986dcs}, TRISTAN~\cite{AMY:1995idu,AMY:1997hbe,TOPAZ:1994xqa}), giving in total 99 points.
The calculations are performed to the NLO accuracy and use the corresponding VALO1.0 photon PDFs in the DIS$_{\gamma}$ scheme. 
The
shaded bands represent 
the propagated uncertainties quantified using the \SI{68}{\percent} CI.
For each experimental data point, the error bars include the systematic and statistical uncertainties added in quadrature.
For better legibility, we offset the theory curves and the experimental data by a factor of $0.2k$, where $k=0-5$, depending on the data set.

One can see from \cref{fig:datatheory} that our 
calculations
reproduce very well the data used in the fit, both the central values of the experimental points and the overall
trend of the $x$ dependence.
In the ALEPH 1999, OPAL 1997 and TRISTAN 1994-1997 cases, the description becomes somewhat worse in the last bins corresponding to large $x$, $x > 0.7$, 
where $F_2^{\gamma}$ begins to rise steeply with an increase of $x$.
(To assist in 
visually
combining the TRISTAN data points into groups corresponding to different $Q^2$, one should relate them to their theoretically-predicted counterparts shown by colored filled squares.)

%% file: sec_results_ThePDFs.tex
\begin{figure}[t!]
\centering
\includegraphics[scale=0.6]{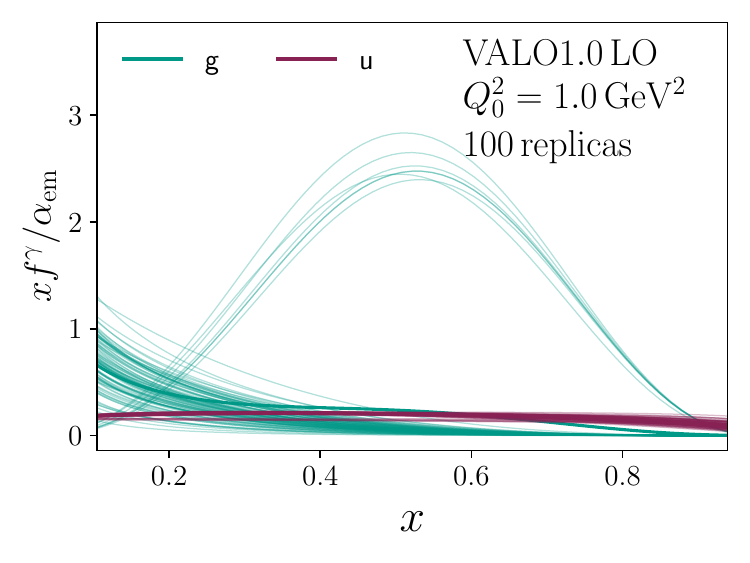}%
\includegraphics[scale=0.6]{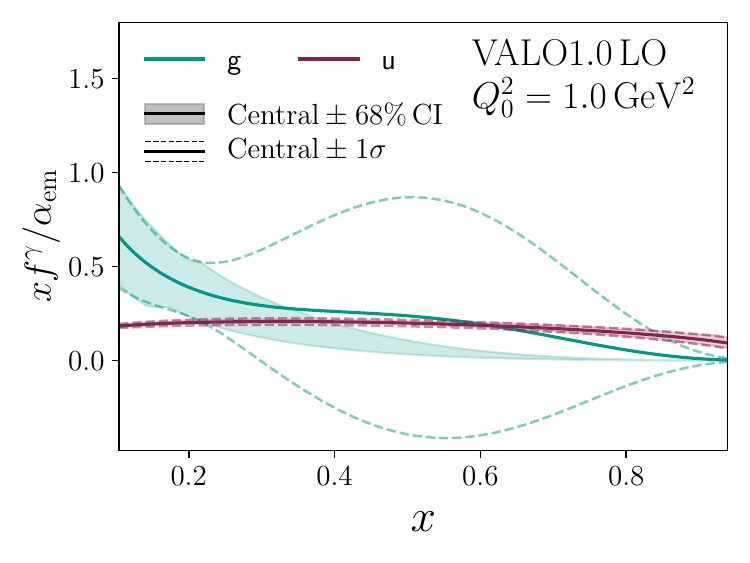}
\includegraphics[scale=0.6]{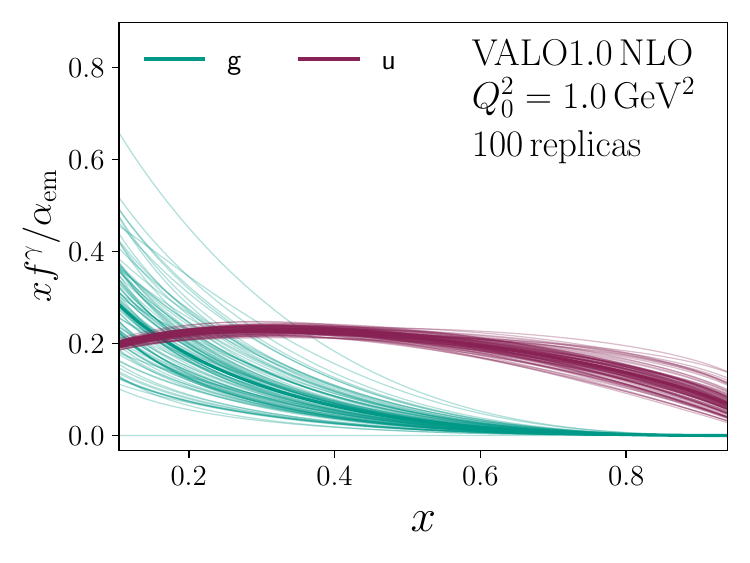}%
\includegraphics[scale=0.6]{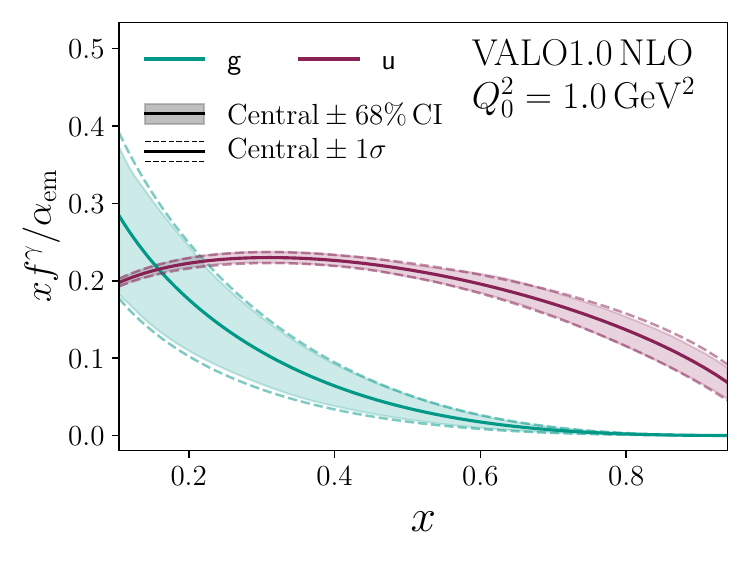}
\caption{
VALO1.0 photon PDFs as a function of $x$ at the initial scale $Q_0^2=1$ GeV$^2$: the rescaled up quark $x u^{\gamma}(x,Q_0^2)/\alpha_{\rm em}$ and the gluon $xg^{\gamma}(x,Q_0^2)/\alpha_{\rm em}$
distributions at LO (upper panels) and NLO (lower panels) as Monte Carlo replicas (left panels) and as central PDFs with associated uncertainty bands (right panels).}
\label{fig:LOQ0}
\end{figure}

\Cref{fig:LOQ0} presents 
VALO1.0 photon PDFs for the up quark and gluon distributions, $x u^{\gamma}(x,Q_0^2)/\alphaem$ and $xg^{\gamma}(x,Q_0^2)/\alphaem$, as a function of the momentum fraction $x$ at the input scale $Q_0^2=\SI{1}{\GeV^2}$.
The upper and lower rows of the panels show the LO and NLO results, respectively.
The left and right panels demonstrate two possible ways to display our results: the left panels give the MC replicas obtained directly from our fitting procedure, while the right panels give the central PDFs as the solid curves, which are calculated by averaging over the replicas, \cref{eq:central}.
The shaded uncertainty bands are calculated using the \SI{68}{\percent} confidence interval (CI), \cref{eq:CI}, whereas the dashed curves give the standard deviation, \cref{eq:st_div}.
Note that that since $d^{\gamma}(x,Q_0^2)=u^{\gamma}(x,Q_0^2)$ and $s^{\gamma}(x,Q_0^2)=K_s u^{\gamma}(x,Q_0^2)$ at the input scale, \cref{eq:input_param}, the down 
and strange quark distributions can be readily read off \cref{fig:LOQ0}.

As can be seen from \cref{fig:LOQ0}, the emerging picture of photon PDFs at the input scale $Q_0^2=\SI{1}{\GeV^2}$ can be summarized as follows.
\begin{itemize}
\item
The quark PDFs are well-constrained and robust: their shapes and magnitudes are similar at LO and NLO. The spread among the replicas is modest, resulting in rather small quark PDF uncertainties across the entire range of $x$.
Moreover, the standard deviation (dotted lines) and the \SI{68}{\percent} CI (shaded bands) uncertainty estimates give similar results indicating a Gaussian distribution of our PDFs.
One notable feature is that $u^{\gamma}(x,Q_0^2)$ decreases faster in the $x \to 1$ limit at NLO than at LO.

\item 
The gluon PDF at NLO is also constrained 
sufficiently
well: 
while the spread of replicas is larger than for the quark distributions,
the replicas do not deviate significantly from the average over the ensemble and the uncertainties are very similar in the 
standard deviation and
\SI{68}{\percent} CI estimates. 
Note that this is partially an artifact of a more rigid parametrization of the gluon distribution at the input scale, \cref{eq:input_param}, which constrains its form for large $x$.  

\item
In contrast, our fit does not reliably constrain the LO gluon distribution.
This can be seen by strong deviations of the MC replicas (upper left panel), which tend to combine in two distinct groups of solutions.
As a result, the uncertainty bands, calculated using the standard deviation and the \SI{68}{\percent} CI, have very different shapes for $x> 0.25$.
The replicas showing a peak structure around $x\approx 0.5$ dominate over all replicas in this region, which yields some peculiar features:
first, they drive the average outside the confidence interval, and, second, they also increase the standard deviation, which eventually makes the symmetric distance to the central PDF even negative for $x > 0.3$  (upper right panel),
even though each replica is positive definite.

\end{itemize}

\begin{figure}[t!]
    \begin{center}
    \includegraphics[width=1.0\linewidth]{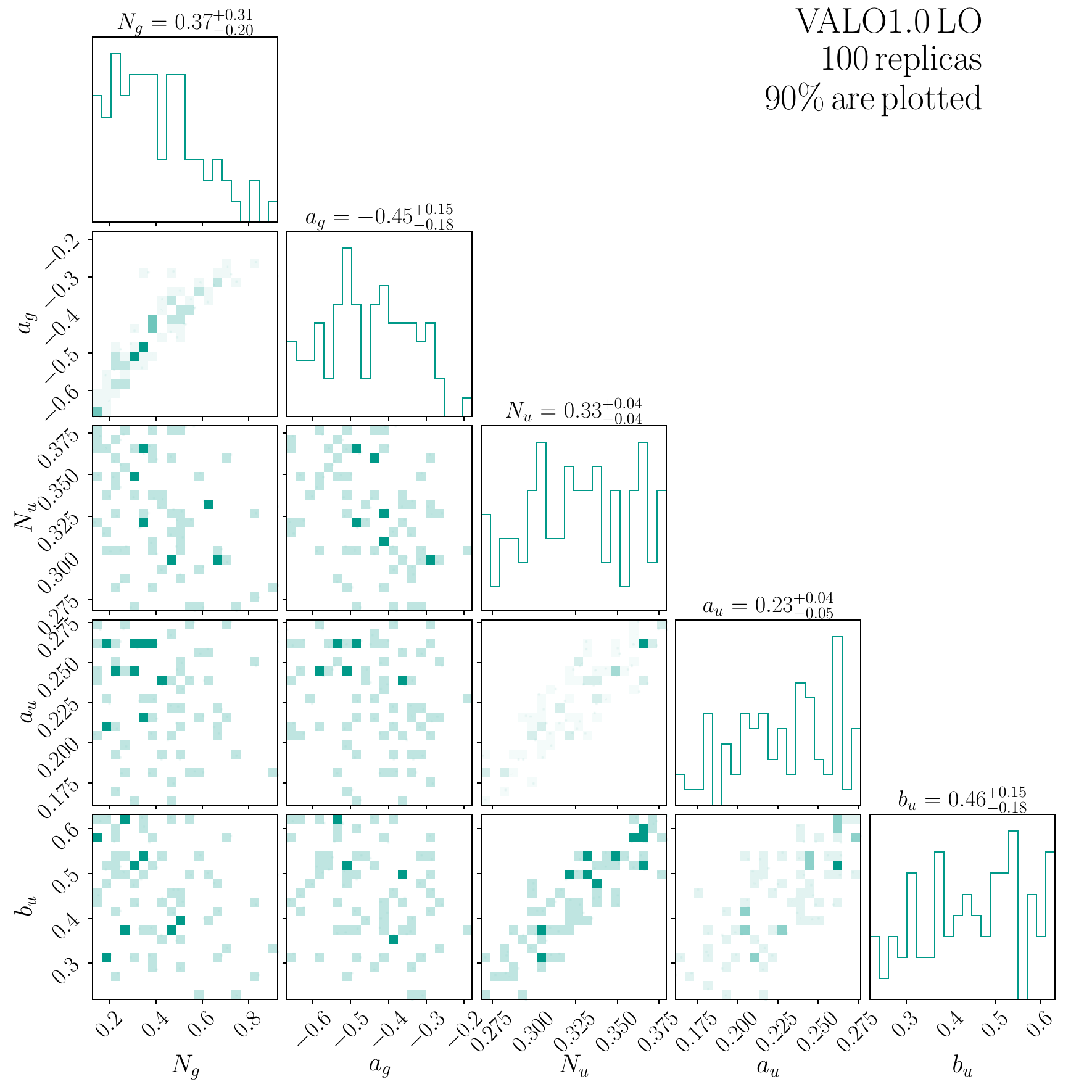} 
    \end{center}
\caption{The median and the \SI{68}{\percent} CI for the fit parameters $(N_g, a_g, N_u, a_u,b_u)$ at LO, calculated using \SI{90}{\percent} of the replicas. }
    \label{fig:params_LO}
\end{figure}

\begin{figure}[t!]
    \begin{center}
    \includegraphics[width=1.0\linewidth]{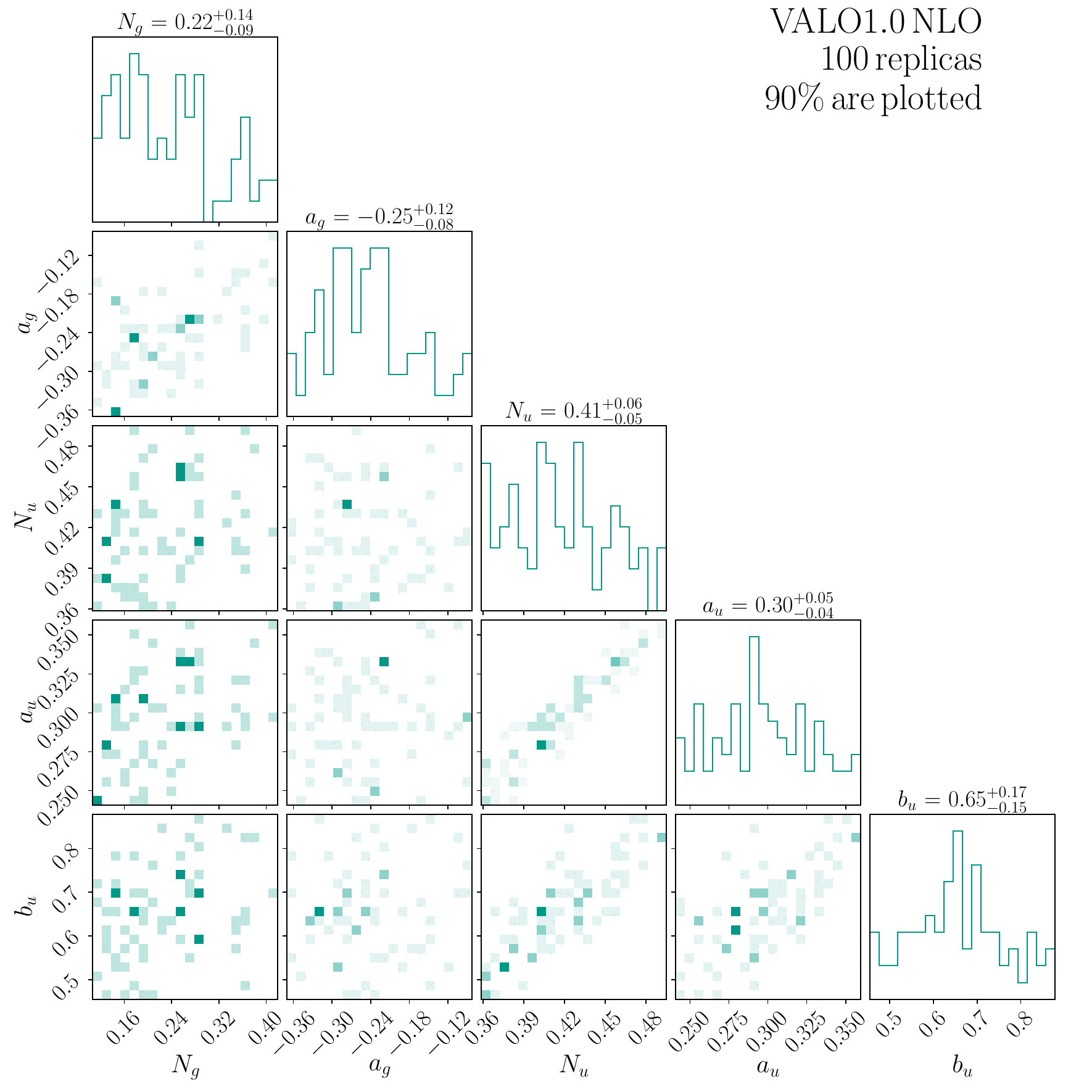}
    \end{center}
    \caption{The $(N_g, a_g, N_u, a_u,b_u)$ fit parameter distributions at NLO, see figure~\ref{fig:params_LO} and text for details.}
    \label{fig:params_NLO}
\end{figure}

Overall, while the fit to $F_2^{\gamma}$ at LO determines only the quark PDFs in the photon and leaves the gluon distribution largely unconstrained,
the fit to $F_2^{\gamma}$ at NLO, which is directly sensitive to gluons,
delivers tightly constrained quark distributions and a reasonably well constrained gluon distribution.
The narrow uncertainty bands 
associated with
the quark distributions reflect the small statistical and systematic errors of the fitted experimental points.

This is further illustrated by \cref{fig:params_LO,fig:params_NLO}, presenting the fit parameters and their correlations at LO and NLO, respectively.
More precisely, the quoted values represent the median and the \SI{68}{\percent} CI for $(N_g, a_g, N_u, a_u,b_u)$.
To improve the readability of the plots, we only show 90 replicas.
One can see from these figures that the fit parameters for the quark 
PDFs $(N_u, a_u,b_u)$ have small uncertainties and are rather close at LO and NLO.
At the same time, the fit parameters for the gluon distribution $(N_g, a_g)$, especially the normalization parameter $N_g$ at LO, are determined with a much lower accuracy and also noticeably differ between the LO and NLO fits.
One should also notice certain correlations among the quark fit parameters at both LO and NLO as well as the correlation between the gluon $N_g$ and $a_g$ parameters at LO.

\begin{figure}[t!]
\begin{center}
    \includegraphics[width=0.5\linewidth]{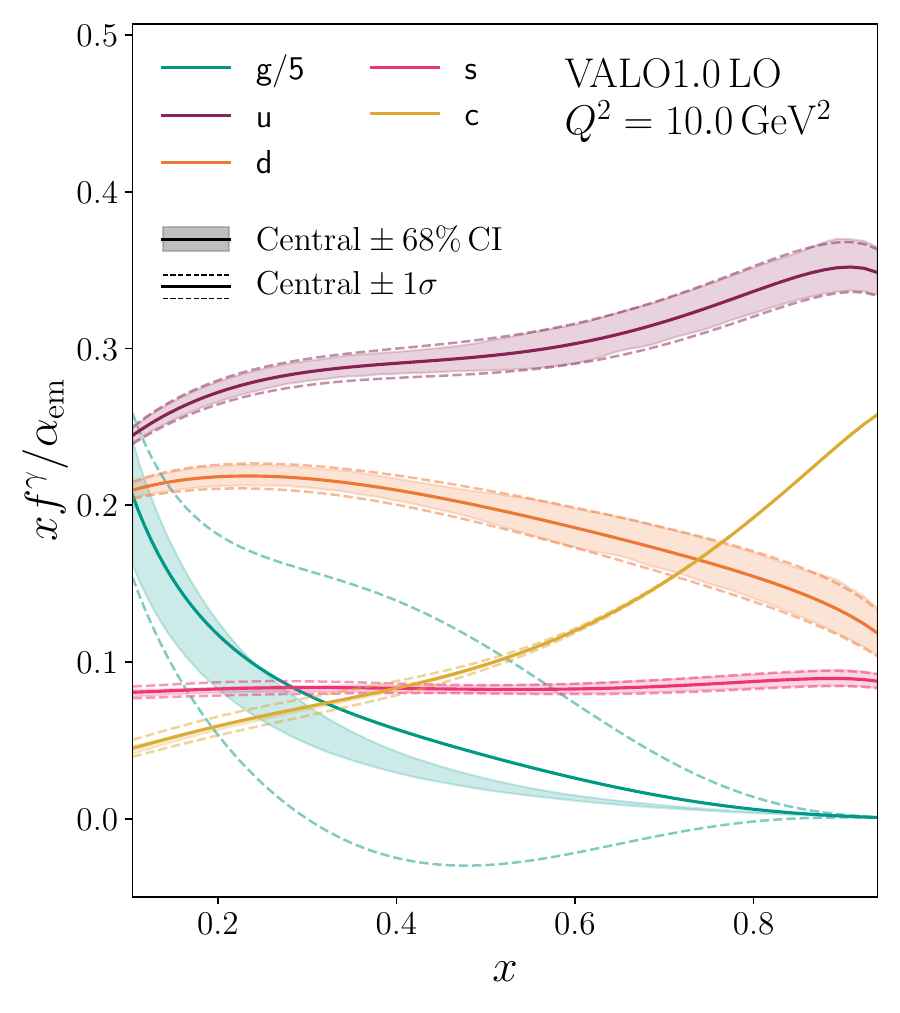}%
    \includegraphics[width=0.5\linewidth]{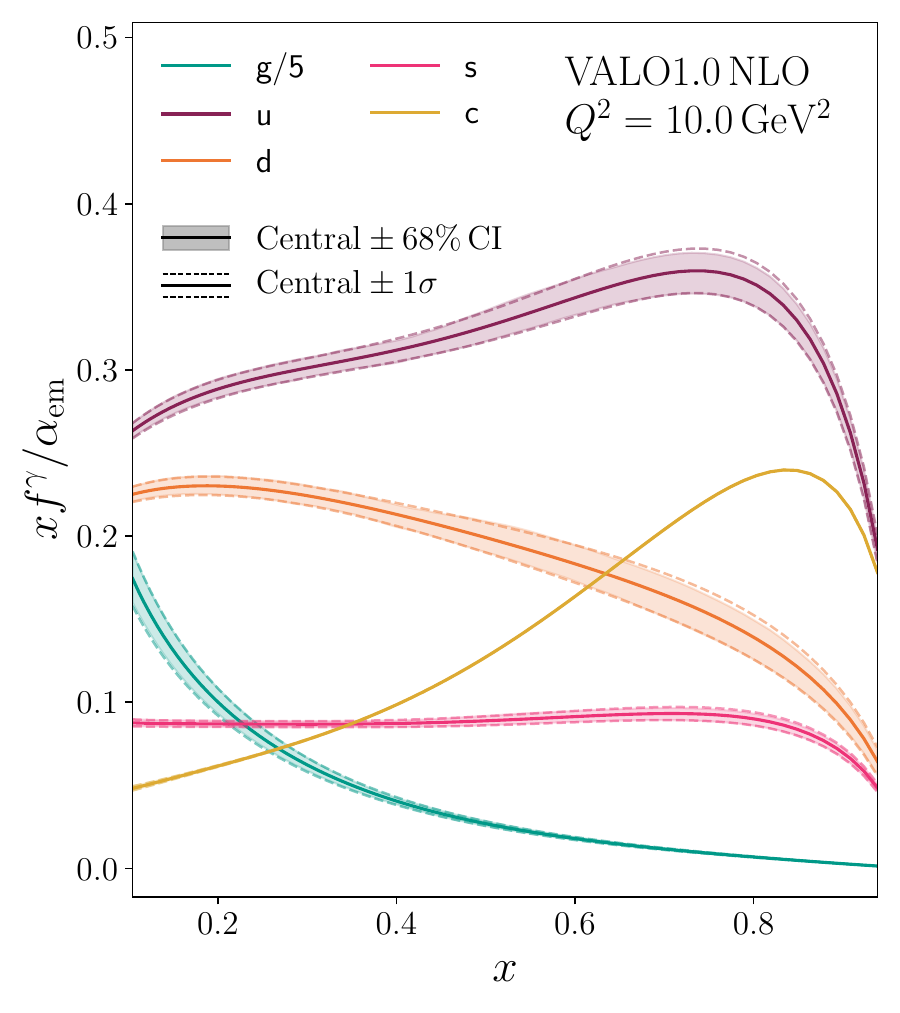}
    \includegraphics[width=0.5\linewidth]{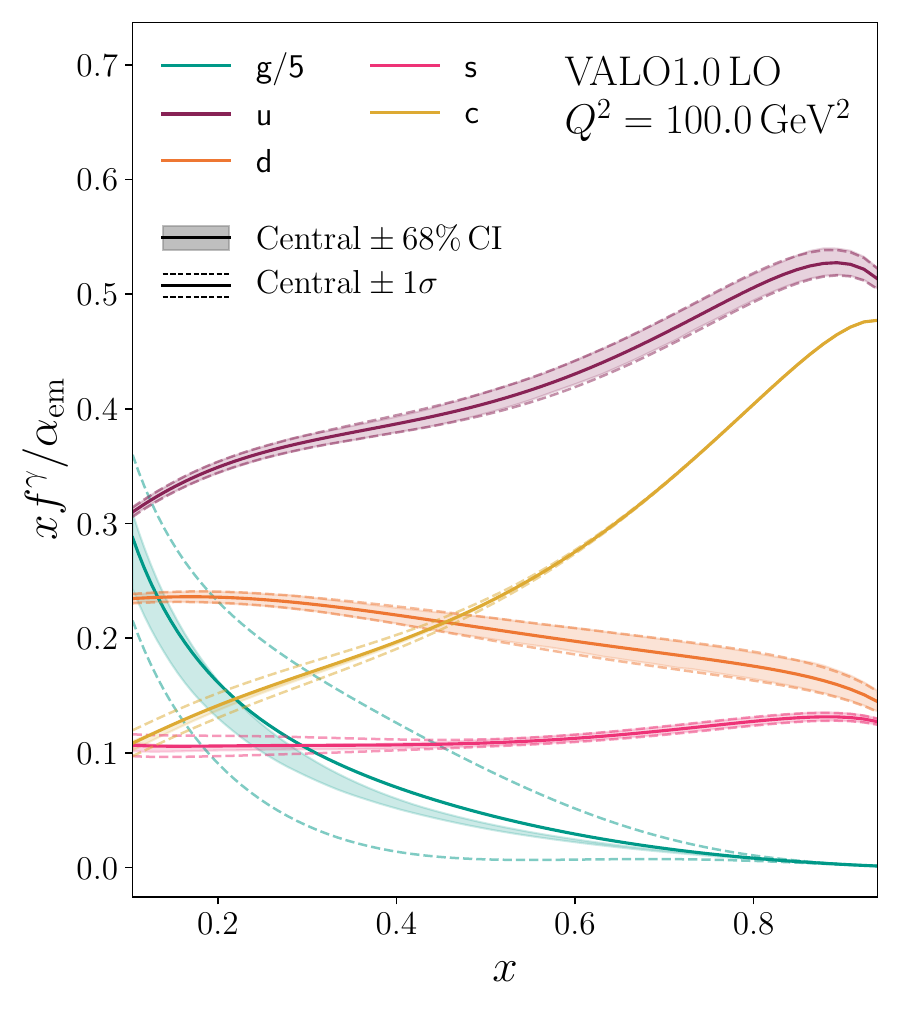}%
    \includegraphics[width=0.5\linewidth]{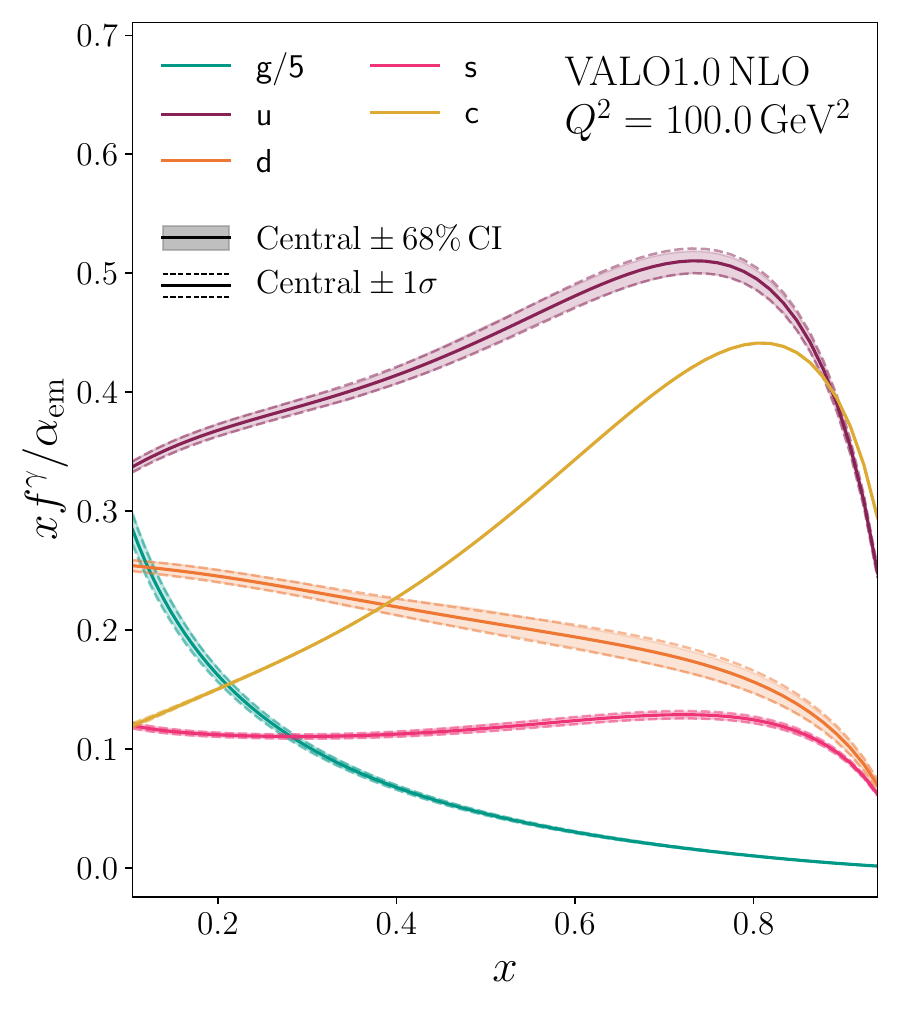}%
    \end{center}
    \caption{Evolved VALO1.0 photon PDFs for individual parton flavors $xf_j^{\gamma}(x,Q^2)/\alpha_{\rm em}$ as a function of $x$ at $Q^2=10$ GeV$^2$ (upper panels) and $Q^2=100$ GeV$^2$ (lower panels): the central PDFs and the uncertainty bands
    (standard deviation and 68\% CI) at LO (left panels) and NLO
    (right panels).
    }
    \label{fig:LONLOhigh}
\end{figure}

Using the VALO1.0 photon PDFs at $Q_0^2=1$ GeV$^2$ as initial conditions, we perform their scale evolution as outlined in \cref{sec:evolution}.
\Cref{fig:LONLOhigh} shows the evolved photon PDFs for individual parton flavors as a function of $x$ at $Q^2=\SI{10}{\GeV^2}$ (upper panels) and $Q^2=\SI{100}{\GeV^2}$ (lower panels); the left and right panels correspond to the LO and NLO results, respectively.
Note that as in \cref{fig:LOQ0}, we show here the appropriately rescaled photon PDFs $xf_j^{\gamma}(x,Q^2)/\alpha_{\rm em}$. The results presented in \cref{fig:LONLOhigh} can be summarized as follows.
\begin{itemize}
\item
In the shown range of $x$, the scale evolution of the quark distributions is rather weak.
The most notable feature is the difference between the up-type quark ($u^\gamma,c^\gamma$) and the down-type quark ($d^\gamma,s^\gamma$) distributions for large $x$, which is driven by the inhomogeneous term in the scale evolution equations.
Indeed, since this term is proportional to squares of quark electric charges $e^2_{q_j}$, see \cref{eq:k_lo}, it is more prominent for the up-type quarks than for the down-type quarks.

\begin{figure}[t!]
\begin{center}
    \includegraphics[scale=0.47]{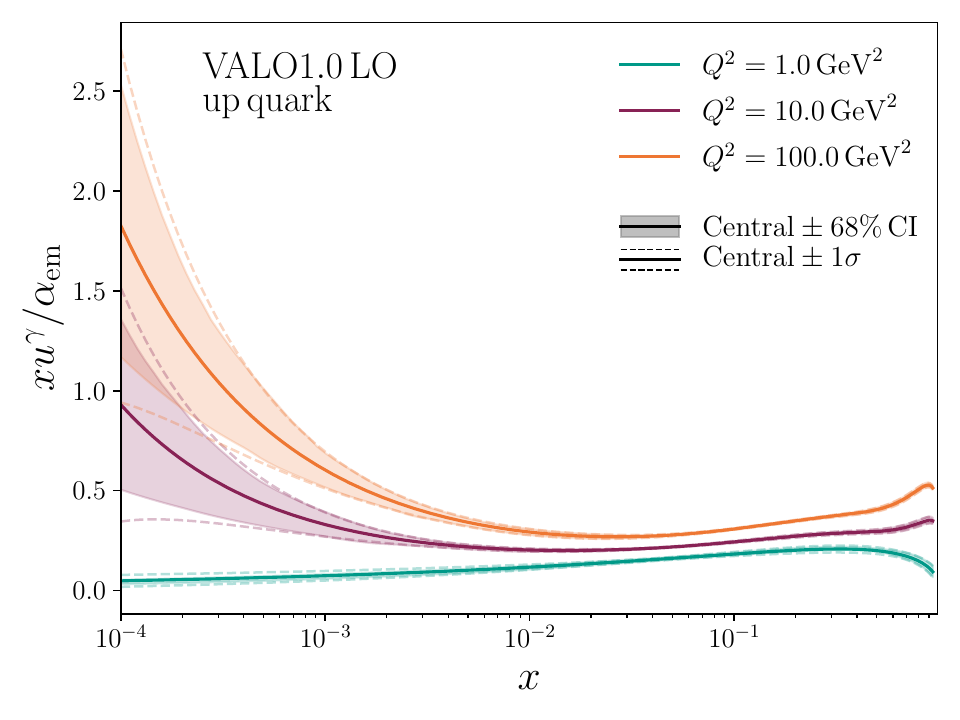}%
    \includegraphics[scale=0.47]{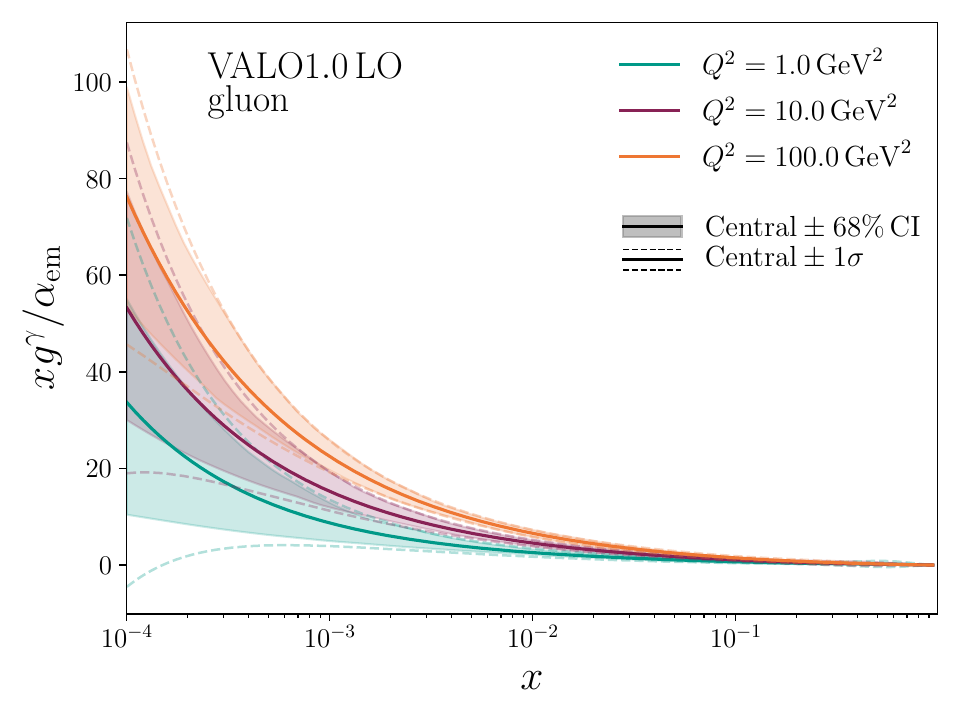}
    \includegraphics[scale=0.47]{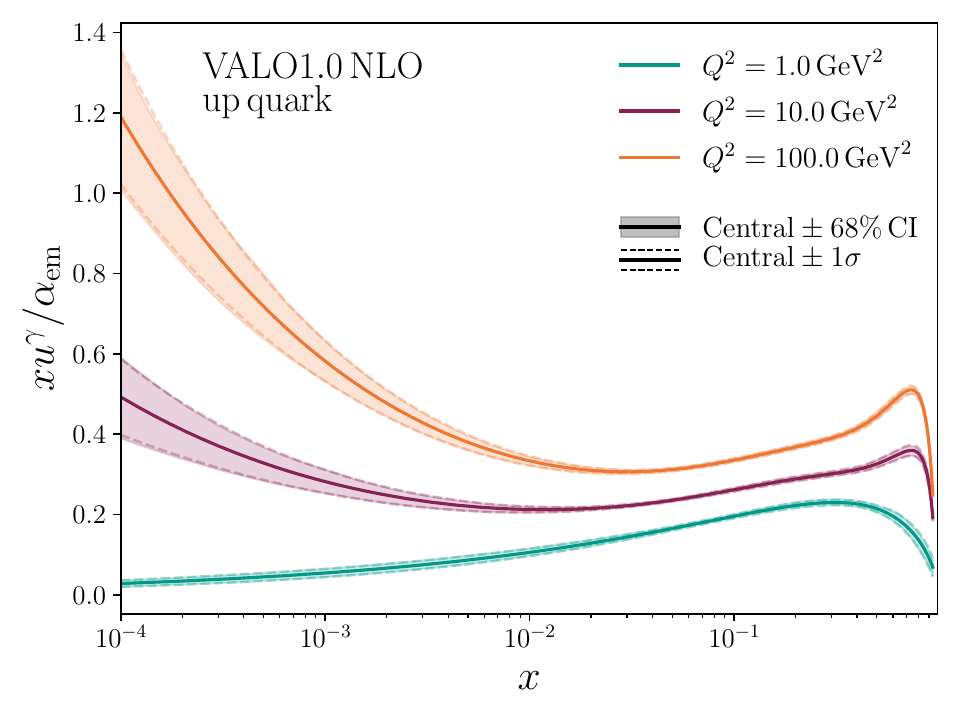}%
    \includegraphics[scale=0.47]{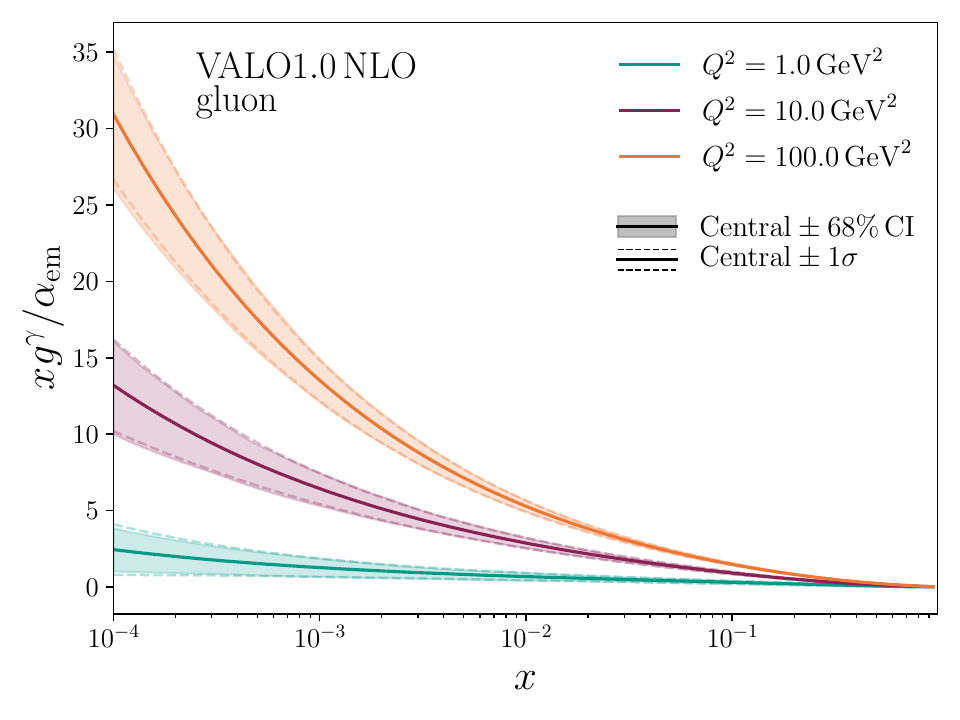}%
    \end{center}
    \caption{Small-$x$ behavior of VALO1.0 photon PDFs: the LO (upper panels) and NLO
    (lower panels) up quark (left panels) and gluon (right panels) distributions as a function of $x$ at $Q^2=(1,10,100)$ GeV$^2$.}
    \label{fig:LONLOhighlog}
\end{figure}

\item
Another feature, which is also related to the inhomogeneous term in scale evolution of photon PDFs, is the observation that despite the fact that $f^{\gamma}(x=1,Q_0^2)=0$, the PDFs do not vanish in the $x \to 1$ limit after the evolution.
Moreover, at NLO, the quark distributions rapidly fall and become negative for very large $x>0.95$.
While this region is at the border of the $x\text{-}Q^2$ kinematic coverage by the $F_2^{\gamma}$ world data, see \cref{fig:kin}, and does not affect phenomenological applications of our photon PDFs, it 
appears to be a genuine feature that has been neglected in the literature.
For instance, the photon PDFs at very large $x$ in ref.~\cite{Slominski:2005bw} are simply forced to vanish, see also the discussion in \cref{sec:PDFcomparisons}.

\item
The scale evolution of the gluon distribution resembles that one in the proton case, with a characteristic rapid rise for small $x$ due to the $g \to gg$ and $q \to gq$ DGLAP splitting
functions.

\item
For the quark and gluon distributions, the uncertainty bands are rather modest, which reflects the propagation of small uncertainties of our photon PDFs at the initial scale. The relative uncertainties for up-type quarks are in general smaller than those of the down-type quarks due to the more prominent contribution from the anomalous component in the evolution equations, as discussed above. Also, as in \cref{fig:LOQ0}, the standard deviation uncertainty band for the LO gluon distribution
deviate from the \SI{68}{\percent} CI curves given by the dotted lines.

\end{itemize}

\Cref{fig:LONLOhighlog} emphasizes the low-$x$ behavior of VALO1.0 photon PDFs.
It shows the LO and NLO up quark and gluon distributions as a function of $x$ on the logarithmic $x$-scale at the input scale $Q_0^2=1$ GeV$^2$ and after the scale evolution 
up to
$Q^2=(10, 100)$ GeV$^2$. One can see that the central values of quark and gluon distributions increase with decreasing $x$ due to the scale-evolution effects.
While the absolute uncertainty bands become wider, the relative uncertainties decrease since they are dominated by evolution effects~\cite{Ball:1994du}.

\begin{figure}[h]
        \centering
        \includegraphics[scale=0.47]{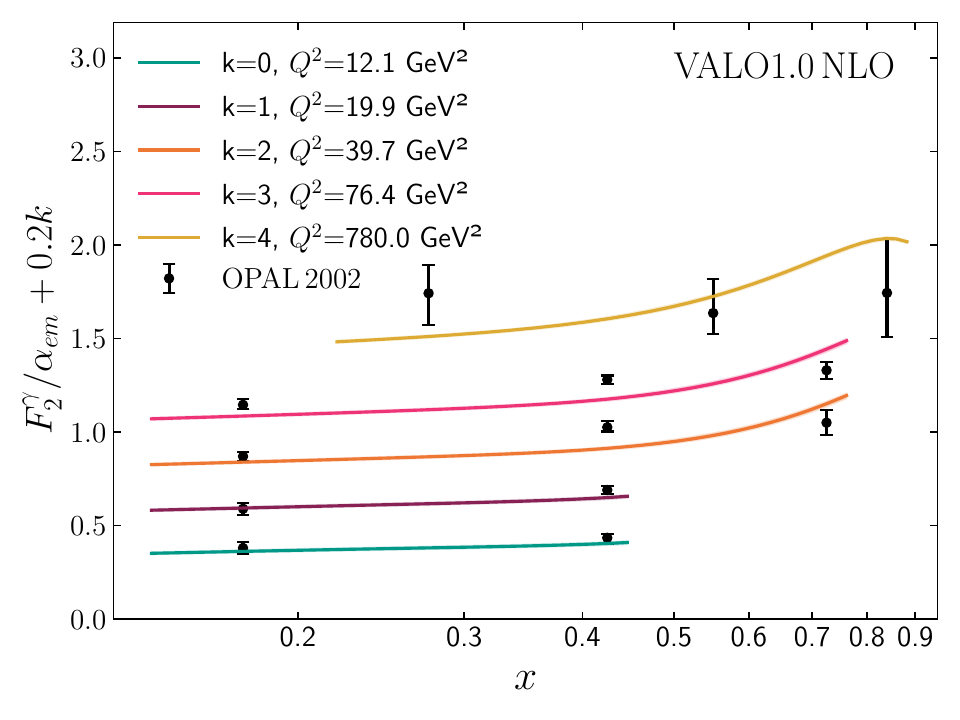}%
        \includegraphics[scale=0.47]{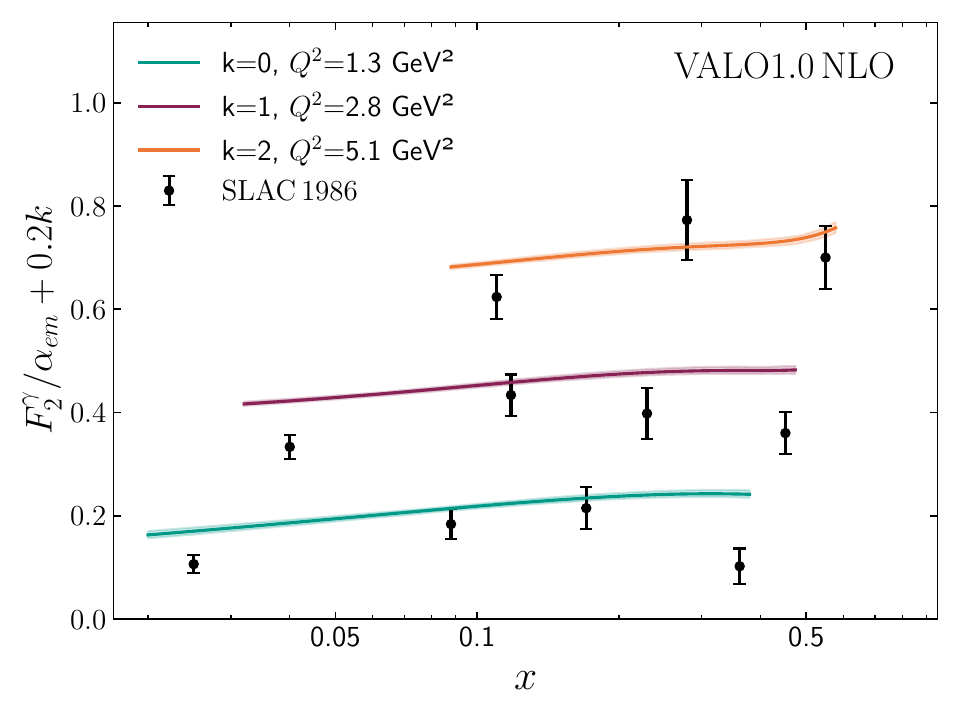}
    \caption{VALO1.0-based predictions for $F_2^{\gamma}$ at NLO 
    as a function of $x$ at fixed values of $Q^2$ and comparison with 
    the OPAL 2002 \cite{OPAL:2002vci} and SLAC 1986 \cite{TPCTwoGamma:1986ycp} data, excluded from our global fit. 
    }
    \label{fig:pheno}
\end{figure}

As a phenomenological application of our 
photon PDFs, in \cref{fig:pheno},
we compare VALO1.0-based predictions for the photon structure function $F_2^{\gamma}$ at NLO with the OPAL 2002 \cite{OPAL:2002vci} and SLAC 1986 \cite{TPCTwoGamma:1986ycp} data, excluded from our global fit. As in \cref{fig:datatheory}, we introduced the offset factor
of $0.2k$ with $0\leq k \leq 5$ for legibility.
One can see from the figure that while our calculation describes the overall shape of $F_2^{\gamma}$ as a function of $x$, it does not reproduce the absolute values of $F_2^{\gamma}$, which are reported with very small experimental uncertainties.
This supports our decision to not include these data in our fit.
Note that we do not show SLAC data points for $Q^2<\SI{1}{\GeV^2}$ to avoid extrapolation beyond our LHAPDF grids.

%% file: sec_results_PDFcomparisons.tex
In this section, we compare our VALO1.0 photon PDFs with selected photon PDFs available in the literature.
These include GRVLO~\cite{Gluck:1991jc} (Glück-Reya-Vogt), CJKL~\cite{Cornet:2002iy} (Cornet-Jankowski-Krawczyk-Lorca), and SaSG~\cite{Schuler:1995fk} (Schuler-Sjöstrand) at LO and GRVHO~\cite{Gluck:1991jc}, CJK~\cite{Cornet:2004nb} (Cornet-Jankowski-Krawczyk), and SAL~\cite{Slominski:2005bw} (Slominski-Abramowicz-Levy) at NLO.
We also highlight some key features of these parameterizations; for a review of photon PDFs available prior to the year 1999, see ref.~\cite{Nisius:1999cv}.

\begin{figure}
    \centering
        \includegraphics[scale=0.5]{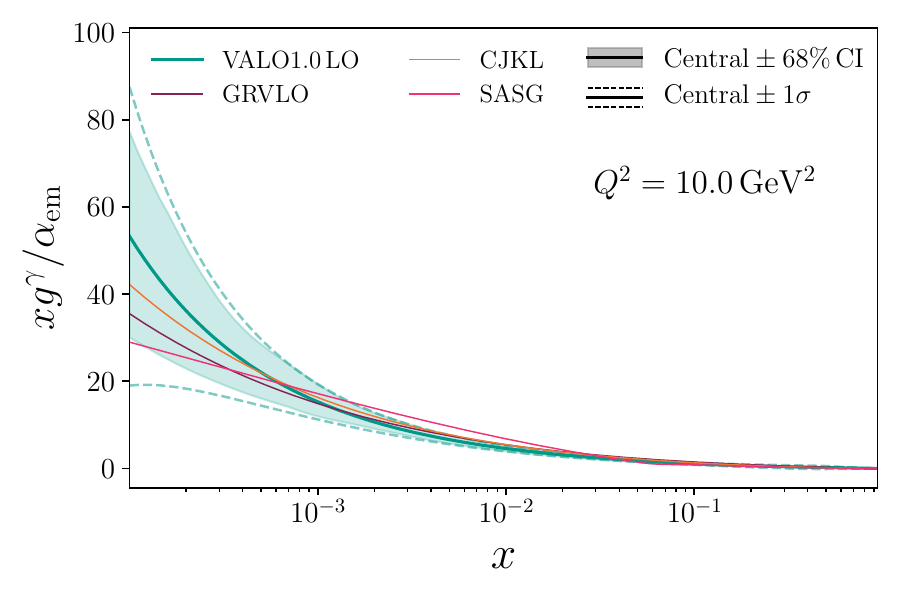}%
    \includegraphics[scale=0.5]{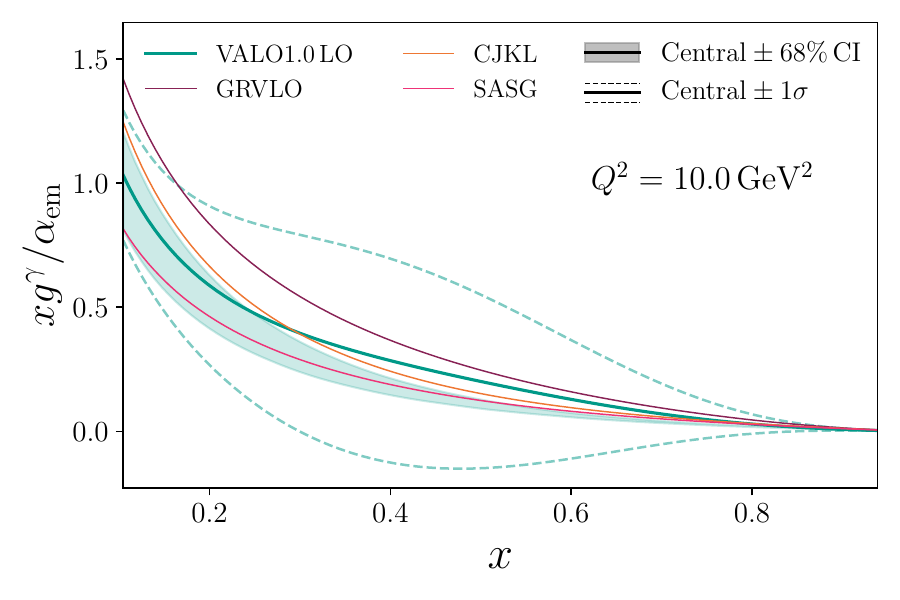}    
        \includegraphics[scale=0.5]{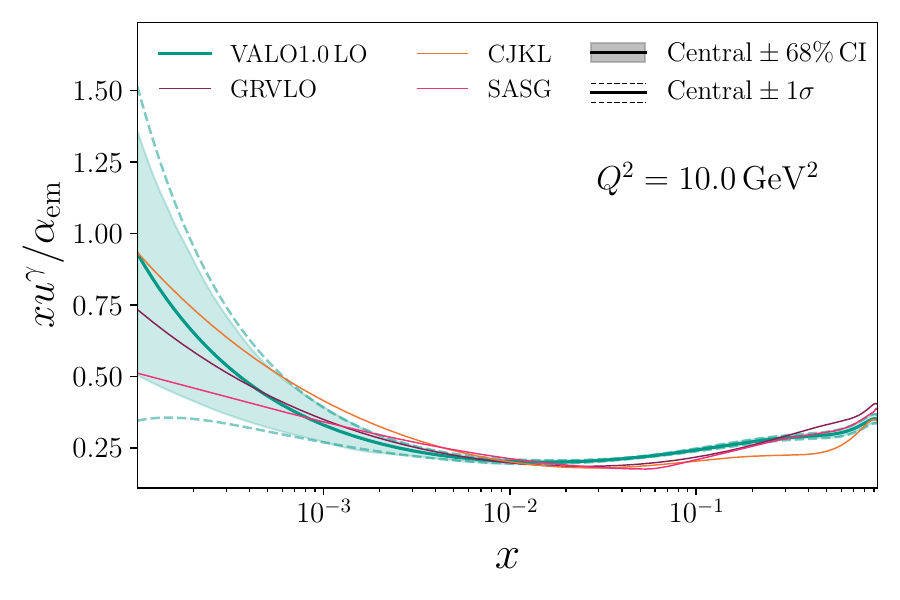}%
        \includegraphics[scale=0.5]{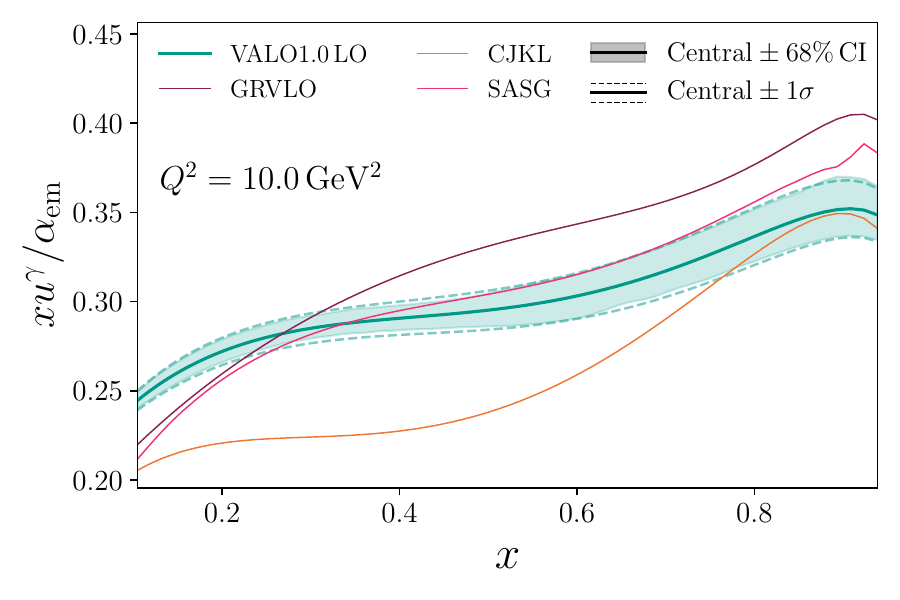}
    \caption{Comparison of VALO1.0 LO photon PDFs with GRVLO~\cite{Gluck:1991jc}, CJKL~\cite{Cornet:2002iy}, and  SASG~\cite{Schuler:1995fk} PDFs:
    the gluon (upper panels) and up quark (lower panels) distributions, $g^{\gamma}(x,Q^2)/\alpha_{\rm em}$
and $u^{\gamma}(x,Q^2)/\alpha_{\rm em}$, as a function of $x$ at $Q^2=10$ GeV$^2$.
The left and right panels use the logarithmic and linear scales for the $x$-axis, respectively.}
    \label{fig:referancePDFsLO}
\end{figure}

\begin{figure}
    \centering
    \includegraphics[scale=0.5]{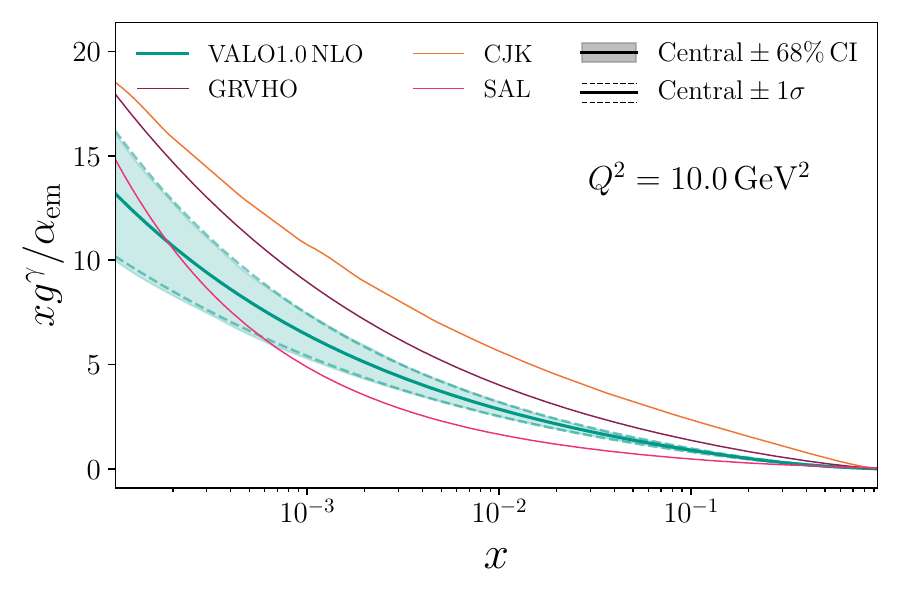}%
    \includegraphics[scale=0.5]{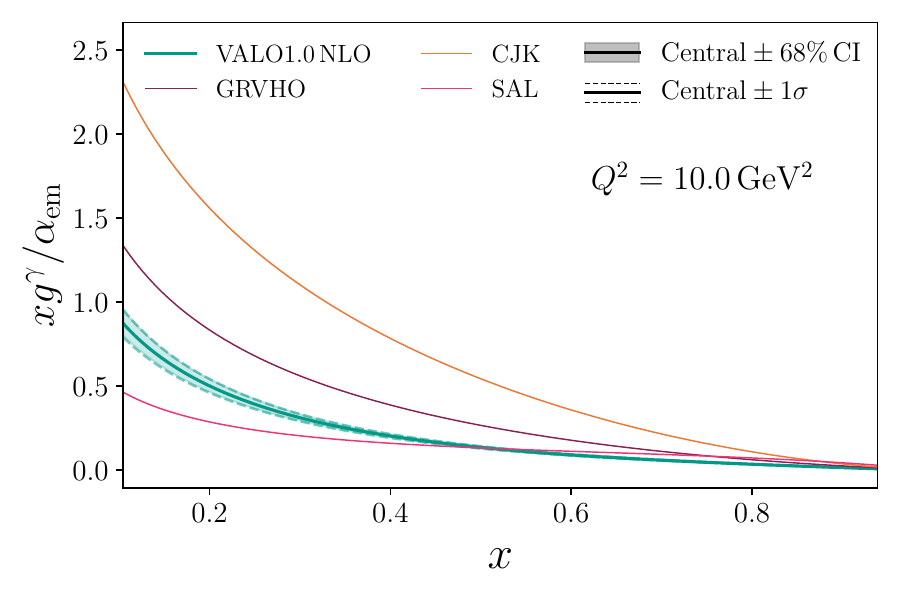}
    \includegraphics[scale=0.5]{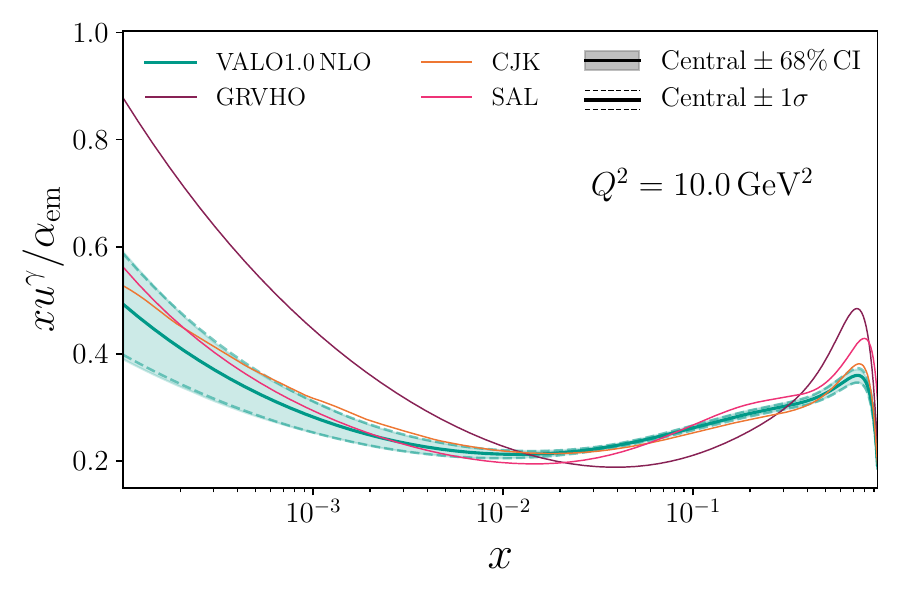}%
    \includegraphics[scale=0.5]{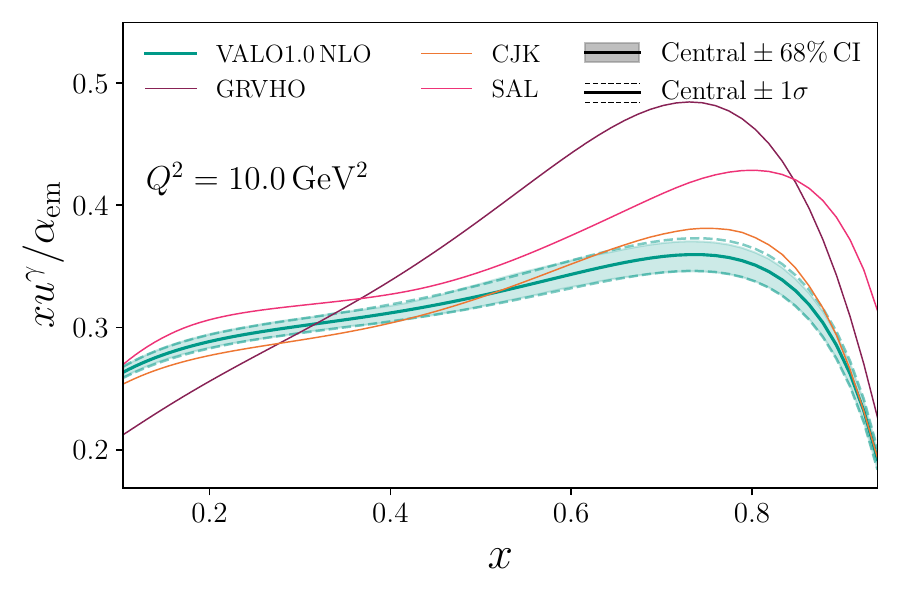}
    \caption{Comparison of VALO1.0 NLO photon PDFs with GRVHO~\cite{Gluck:1991jc},
    CJK~\cite{Cornet:2004nb}, and SAL~\cite{Slominski:2005bw} PDFs at $Q^2=10$ GeV$^2$; see the caption of \cref{fig:referancePDFsLO} for details.
    }
    \label{fig:referancePDFsNLO}
\end{figure}

\Cref{fig:referancePDFsLO,fig:referancePDFsNLO} show the gluon and up quark distributions, $xg^{\gamma}(x,Q^2)/\alpha_{\rm em}$ and $xu^{\gamma}(x,Q^2)/\alpha_{\rm em}$, as a function of $x$ at $Q^2=10$ GeV$^2$ at LO and NLO, respectively.
Different curves correspond to the photon PDF parametrizations mentioned above.
To make the comparison more comprehensive, each panel with the logarithmic $x$-scale (left panels) has its counterpart with the linear $x$-scale (right panels). One can see from these figures that the shapes of the LO and NLO VALO1.0 photon PDFs broadly agree with the selected PDFs available in the literature. 
The biggest numerical differences can be found in the quark and gluon distributions in the small $x$ limit: while our uncertainty bands reasonably cover the spectrum of the other fits at LO, the narrower uncertainty bands at NLO 
make the differences more pronounced.

The GRVLO and GRVHO \cite{Gluck:1991jc} PDFs correspond to three light
active quark flavors ($u$, $d$, $s)$, 
implementing charm and bottom quarks via the Bethe-Heitler (BH) formula and 
treating them as massless at high values of the photon-photon energy $W^2=Q^2(1/x-1)$ in scale evolution (only the data with $W>\SI{2}{\GeV}$ 
was
used).
The input distributions are taken in the purely hadron-like form
motivated by %
VMD
arguments,
\begin{equation}
    f_j^{\gamma}(x,Q_0^2) = \kappa \frac{4\pi \alpha_{\text{em}}}{f_\rho^2} f_{j}^{\pi}(x,Q_0^2) \,,
    \label{eq:grv_input}
\end{equation}
where 
$f_{j}^{\pi}(x,Q_0^2)$ 
are the pion PDFs, $f_{\rho}$ is the photon-meson coupling constant, and 
$\kappa$ is the only free fit 
parameter.
We note that by the time of the GRV analysis, the LEP data did not yet exist, which explains the significant deviations between GRV and the other determinations.

The two LO SaSG sets~\cite{Schuler:1995fk}, SaS1 using $Q^2_0=0.36$ GeV$^2$ as a starting scale and SaS2 with $Q^2_0 = 4$ GeV$^2$, employ the massless evolution equations for light quarks and the BH expression for heavy quarks.
The PDFs at the initial scales are parameterized in the VMD-inspired form, with the main difference between SaS1 and SaS2 being in accounting 
for
the contribution of higher mass vector mesons in the latter case.
The parametrizations are provided in the $\text{DIS}_\gamma$ and $\overline{\text{MS}}$ factorization schemes, where the scheme dependence originates from including the $C_\gamma^{(1)}(x)$ coefficient defined in \cref{sec:coeff_fnc} into their $\overline{\text{MS}}$ fit already at LO.
The momentum sum rule 
is implemented for the hadronic component of photon PDFs.

The SAL PDFs~\cite{Slominski:2005bw} are based on the input parametrization, which is given by a sum of point-like and hadron-like contributions.
In addition to $F_2^{\gamma}$ data, this analysis also includes data on the $F_2^{p}$ structure function and the cross section of dijet photoproduction in electron-proton scattering at HERA (the former is done in a model-dependent way using the so-called Gribov factorization relation).
The CJK~\cite{Cornet:2004nb} and CJKL~\cite{Cornet:2002iy} PDFs parametrize their input distributions using a hadron-like form motivated by the VMD model,
\begin{equation}
    f_j^{\gamma}(x,Q_0^2) = \kappa \frac{4\pi \alpha_{\text{em}}}{f_\rho^2} f_{j}^{\rho}(x,Q_0^2) \,,
    \label{eq:cjk_input}
\end{equation}
where $f_{j}^{\rho}(x,Q_0^2)$ are the $\rho$ meson PDFs, and 
$\kappa$
accounts for contributions of other vector mesons.
The distributions $f_{j}^{\rho}(x,Q_0^2)$ are parametrized in a hadron-like form
$N x^{\alpha}(1-x)^{\beta}$, where
$N$ for quarks and gluons
are constrained by the valence number and the energy-momentum sum rules, which leaves $\kappa,\alpha,\beta$ as free parameters.

%% file: sec_summary.tex
In this work we perform a global QCD analysis of the world data on the photon structure function $F_2^{\gamma}$ measured in $e^{+} e^{-}$ scattering
and determine new sets of LO and NLO photon PDFs with uncertainties, which we refer to as VALO1.0 PDFs. Our statistical analysis of the $F_2^{\gamma}$ data is based on 100 Monte Carlo (MC) replicas, which are used to calculate the central PDFs as average over the replicas and to estimate PDF uncertainties, using the 68\% CI as well as the standard deviation.

We specify the boundary conditions for our PDFs at the input scale in terms of 5 free parameters: 3 for quarks, with assumed isospin symmetry for the up and down quarks and a VMD-motivated suppression for the strange quarks, and 2 for gluons, with the large-$x$ behavior fixed by the quark counting rules.
This yields a good convergence of both LO and NLO fits.
The resulting quark PDFs are well-constrained and robust, with similar shapes and fit parameters at LO and NLO, and are characterized by a small spread of MC replicas from the central value.
Also the gluon distribution at NLO is determined reasonably well within our restricted ansatz.
On the other side, the fitted data do not reliably constrain the gluon distribution at LO, resulting in strong deviations of the MC replicas from the average and leading to mutually different shapes for the standard deviation and 68\% CI uncertainties.
Overall, the rather precise $F_2^{\gamma}$ data 
and our restricted parametrization allow only for modest uncertainties for the quark and gluon distributions for $x > 0.01$, which increase for $x<0.01$ where the data constraints become limited.

Our photon PDFs broadly agree with the parameterizations of photon PDFs available in the literature, with the largest differences found in the small-$x$ behavior of the quark and gluon distributions, where 
the data are sparse
and
the PDF uncertainties are large. In the future, they can be better constrained by including in a global QCD analysis the data on dijet photoproduction in electron-proton scattering at HERA
and provide an updated baseline for photoproduction studies in UPC at the LHC and at the EIC.

We provide LO and NLO VALO1.0 photon PDFs in the form of LHAPDF6 grids for
the central PDFs and
100 MC replicas, 
both in the DIS$_{\gamma}$ and $\overline{\rm MS}$ factorization schemes.
Other important deliverables of this work include the code $\gamma$\texttt{EKO}, which 
solves
the inhomogeneous scale evolution equations for photon PDFs in Mellin space,
and the numerical fitting framework \texttt{VALOfitter}; both of them are available 
as open-source tools.

%% file: app_splitting_functions.tex
In this appendix, we summarize the explicit 
expressions for the coefficient functions 
and the splitting functions, which we used in our work.

In the $\overline{\rm MS}$ factorization scheme,
the photon structure function
$F_2^{\gamma}$ in \cref{eq:F2_MSbar} involves the 
NLO quark and gluon coefficient functions, 
$C_q^{(1)}(z)$ and $C_g^{(1)}(z)$, and the NLO photon coefficient function
$C_{\gamma}^{(1)}(z)$.
They are given by the following standard expressions~\cite{Gluck:1983bh}
\begin{align}
C_q^{(1)}(z) &= C_F\Bigg[\frac{9+5z}{2}-2 \frac{1+z^2}{1-z}\ln z-\frac{3}{2} \frac{1+z^2}{(1-z)_{+}}
+ 2 (1+z^2) \left(\frac{\ln(1-z)}{1-z}\right)_{+} \nonumber \\
& \hspace{40pt} -\left(9+\frac{2 \pi^2}{3}\right)\delta(1-z) \Bigg] \,, \nonumber\\
C_g^{(1)}(z) &=4 T_R \Bigg[\left(z^2+(1-z)^2\right) \ln\left(\frac{1-z}{z}\right)-1+8z(1-z) \Bigg]\,, \nonumber \\
C_{\gamma}^{(1)}(z) &= 2 N_c\, C_g^{(1)}(z) \,,
\end{align}
where $C_F=4/3$, $T_R=1/2$, $N_c=3$ is the nuber of colors, and $(\dots)_{+}$ are the plus-distributions.

To evaluate $F_2^{\gamma}$ in the DIS$_{\gamma}$ factorization scheme,
one uses the same quark and gluon coefficient functions and sets the photon
coefficient function to zero, see \cref{eq:F2_DIS_gamma}.

The scale evolution of photon PDFs is governed by eqs.~(\ref{eq:ev_matrix})--(\ref{eq:k_expansion}), where $P_{ij}^{(0,1)}$ are the standard DGLAP 
parton-parton splitting functions, and $k^{(0,1)}$ are the splitting functions due to the $\gamma \to q {\bar q}$ point-like coupling. The latter give rise to the inhomegeneous terms and make the scale evolution of photon PDFs different from that in the proton case. 

 To LO accuracy, the parton-parton splitting functions read
 \begin{align}
P_{qq}^{(0)}(x) &= C_F \left(\frac{1+x^2}{1-x}\right)_{+}\nonumber\\
P_{qg}^{(0)}(x)&=T_R (x^2+(1-x)^2) \,, \nonumber\\
P_{gq}^{(0)}(x) &= C_F \frac{1+(1-x)^2}{x} \,, \nonumber\\
P_{gg}^{(0)}(x) &= 2 C_A \left(\frac{x}{(1-x)_{+}}+\frac{1-x}{x}+x(1-x)+\left(\frac{11}{12}-\frac{n_f}{18}\right)\delta(1-x)\right) \,,
\label{eq:qcd_g11}
\end{align}
where $T_R=1/2$, $C_A=3$, and $n_f$ is the number of quark flavors.
The expressions for the NLO parton-parton splitting functions $P_{ij}^{(1)}$
are rather lengthy and can be found in ref.~\cite{Gluck:1983mm}.

The LO photonic splitting functions have the following form~\cite{Moch:2001im},
\begin{align}
k^{(0)}(x)& =  x^2+(1-x)^2 \, , \nonumber \\
k_{NS}^{(0)} (x) &= 4 N_c n_de_\Delta^2 \cdot k^{(0)}(x) \, , \nonumber\\
k_{\Sigma}^{(0)} (x)&= 4 N_c e_{\text{tot}}^2 \cdot k^{(0)}(x) \,.
\label{eq:qcd_g12}
\end{align}
In \cref{eq:qcd_g12}, we defined the charge combinations,
\begin{align}
    e_{\text{tot}}^2 = n_u e_u^2 + n_d e_d^2 \qand e^2_\Delta = e_u^2 - e_d^2 \,,
\end{align}
where $n_d$ and $n_u$ refer to the number of up-like (up and charm) and down-like (down, strange and bottom) quarks, respectively. 

At NLO and in the $\overline{\rm MS}$ factorization scheme, the photonic 
splitting functions are given by the following expressions~\cite{Moch:2001im},
\begin{align}
k_{NS}^{(1)} &= 4 N_c n_d e_\Delta^2  k^{(1)}(x) \,, \nonumber\\
k_{\Sigma}^{(1)}&= 4 N_c  e_{\text{tot}}^2   k^{(1)}(x) \,, \nonumber\\
k_g^{(1)} &= 4 N_c e_\text{tot}^2  C_F \left[ -16+8x+\frac{20}{3}x^2+\frac{4}{3x}
-(6+10x)\ln(x)-2(1+x)(\ln(x))^2 \right] \,,
\label{eq:qcd_g13}
\end{align}
where 
\begin{align}
k^{(1)}(x)&=C_F\Big[4-9x-(1-4x) \ln(x)-(1-2x)(\ln(x))^2+4 \ln(1-x)  \nonumber\\
&\hspace{20pt} +(4 \ln(x)-4 \ln(x) \ln(1-x)+2 (\ln(x))^2-4 \ln(1-x)+2 (\ln(1-x))^2 \nonumber\\
&\hspace{20pt} -\frac{2}{3}\pi^2+10) (x^2+(1-x)^2) \Big]\,.
\label{eq:qcd_g14}
\end{align}
The transformation to the DIS$_\gamma$ scheme is given by~\cite{Moch:2001im}
\begin{align}
   k_{NS}^{(1)}(x)_{\text{DIS}_\gamma} &= k_{NS}^{(1)}(x)_{\overline{\rm MS}} \, -  \qty(P_{qq}^{(0)}\otimes C_\gamma^{(1)})(x)\,,\nonumber\\
    k_{\Sigma}^{(1)}(x)_{\text{DIS}_\gamma} &= k_{\Sigma}^{(1)}(x)_{\overline{\rm MS}} \, -  \qty(P_{qq}^{(0)}\otimes C_\gamma^{(1)})(x)\,,\nonumber\\
    k_{g}^{(1)}(x)_{\text{DIS}_\gamma} &= k_{g}^{(1)}(x)_{\overline{\rm MS}} \, -  \qty(P_{gq}^{(0)}\otimes C_\gamma^{(1)})(x) \,.
\end{align}

%% file: app_geko.tex
In this appendix, we discuss the technical details, which are necessary 
to %
perform the scale evolution of photon PDFs.
In particular, we 
present
the algorithm for implementation of the inhomogeneous, point-like 
contribution in the evolution equations in \cref{eq:DGLAP} using
the new open-source code $\gamma$\texttt{EKO} (pronounced [geko]) available from our repository\footnote{\url{https://github.com/felixhekhorn/geko}} with the associated documentation available online\footnote{\url{https://geko.readthedocs.io/}}.
As 
stated in \cref{sec:evolution}, we follow closely the \texttt{EKO}~\cite{barontini_2025_15878535} framework and, thus, also its implementation strategy.

We solve the evolution equations in Mellin space
and deliver all the results in direct $x$-space, by performing as the final step of the algorithm the inverse Mellin transformation.
We use the standard Lagrange interpolation techniques to form a basis in $x$-space.
However, while the hadronic EKO $\mathbf E$ is a matrix in $x$-space (evolution involves a mixing between initial and final momentum fractions), the point-like contributions are vectors in $x$-space, since the hadronic evolution $\mathbf E$ is always applied to the photonic anomalous dimensions $\mathbf k$.
Thus the inversion is just the (numerical) inversion onto a given interpolation point.

We adopt the normalization of ref.~\cite{Moch:2001im} and expand all perturbative ingredients in powers of $a_s = \alpha_s/(4\pi)$ and $a_{em} = \alphaem/(4\pi)$ for the strong and electromagnetic couplings, respectively.
This applies in particular to the perturbative expansion of the QCD beta function $\beta(a_s)$, the hadronic anomalous dimensions $\gamma(a_s)$, and the photonic anomalous dimensions $\mathbf k(a_s)$.
In particular, we define
\begin{equation}
    \beta(a_s) = -\sum_{j=0}\left(a_s\right)^{2+j} \beta_j, \quad
    \gamma(a_s) = \sum_{j=0}\left(a_s\right)^{1+j} \gamma^{(j)}, \quad
    \mathbf k(a_s) = a_{em}\sum_{j=0}\left(a_s\right)^{j} \mathbf k^{(j)} \,.
    \label{eq:pbgk}
\end{equation}
Performing the standard transformation of variables, $\mu_F^2 \to a_s(\mu_F^2)$, and using the definition of the QCD beta function, \cref{eq:DGLAP} can be rewritten in the following equivalent form,
\begin{equation}
    \frac{\diff}{\diff a_s} \tilde{\mathbf{f}}^\gamma(a_s) = -\frac{\gamma(a_s)}{\beta(a_s)} \cdot \tilde{\mathbf{f}}^\gamma(a_s) - \frac{\tilde{\mathbf k}(a_s)} {\beta(a_s)} \,.
\end{equation}
Below we explicitly solve this equation first at LO and then at NLO, which can also be generalized to higher orders.

\subsection{Leading order}
\label{sec:gekolo}

At LO, one keeps exactly one term in each sum in \cref{eq:pbgk}.
In this case a closed form solution exists for the hadronic EKO $\tilde{\mathbf E}$, the point-like contributions $\tilde{\mathbf f}^\gamma_{inhom}$, and, thus, for the photonic EKO $\tilde{\mathbf E}^\gamma$.
In particular, we find the following expressions in the non-singlet sector,
\begin{align}
    \tilde E_{NS}^{(0)}(a_s \leftarrow a_s^0) &= \exp\left(\gamma_{NS}^{(0)}\ln(a_s/a_s^0) / \beta_0\right)\,,\\
    \tilde f_{{\rm inhom},NS}^{\gamma,(0)}(a_s) &= \frac {a_{em}k_{NS}^{(0)}}{(\gamma_{ns}^{(0)}+\beta_0)} \left(\frac{\exp\left(\gamma_{NS}^{(0)}\ln(a_s/a_s^0) / \beta_0\right)}{a_s^0} -\frac 1 {a_s} \right) \,,
\end{align}
and, similarly, in the singlet sector,
\begin{align}
    \tilde{\mathbf E}_S^{(0)}(a_s \leftarrow a_s^0) &= \sum_{\lambda\in\{+,-\}} \mathbf e_\lambda^{(0)} \exp\left(\gamma_{S,\lambda}^{(0)}\ln(a_s/a_s^0) / \beta_0\right)\,,\\
    \tilde {\mathbf f}_{{\rm inhom},S}^{\gamma,(0)}(a_s) &= \sum_{\lambda\in\{+,-\}} \frac {a_{em}}{(\gamma_{S,\lambda}^{(0)}+\beta_0)} \left(\frac{\exp\left(\gamma_{S,\lambda}^{(0)}\ln(a_s/a_s^0) / \beta_0\right)}{a_s^0} -\frac 1 {a_s} \right) \mathbf e_{\lambda}^{(0)} \tilde{\mathbf k}_S^{(0)}\,.
\end{align}
In these equations, $\gamma_{S,\pm}^{(0)}$ are the eigenvalues of the singlet matrix $\gamma_S^{(0)}$, and $\mathbf e_{\pm}^{(0)}$ are the associated projectors for the eigenvalues.
The LO hadronic anomalous dimensions, $\gamma_{NS}^{(0)}$ and $\gamma_S^{(0)}$, can be found in refs.~\cite{Moch:2004pa,Vogt:2004mw}, and the LO photonic anomalous dimensions, $\tilde k_{NS}^{(0)}$ and $\tilde{\mathbf k}_S^{(0)}$, 
can be found in ref.~\cite{Moch:2001im}.
They can also be obtained through Mellin transformation of the corresponding expressions in \cref{app:splitting_functions}.
Note that our results agree with those in eq.~(2.6) of ref.~\cite{Gluck:1983mm} truncated to the LO accuracy.

\begin{figure}[t!]
\centering
\includegraphics[width=\linewidth]{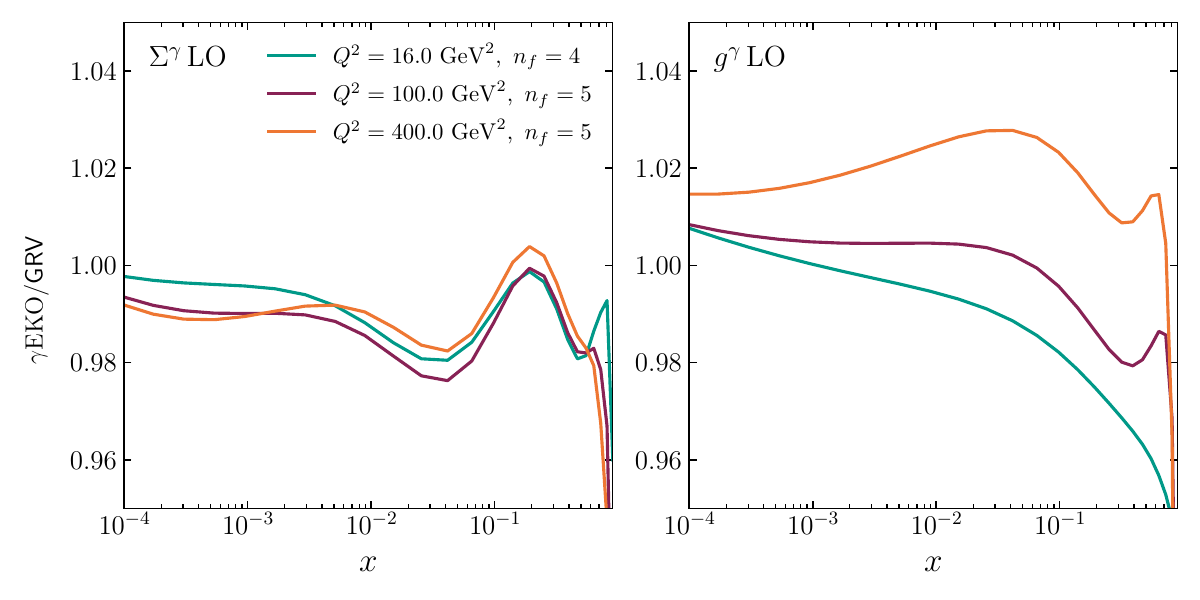}
\caption{Ratios of LO quark singlet $\Sigma^{\gamma}(x,Q^2)$ (left panel) and gluon
$g^{\gamma}(x,Q^2)$ (right panel) distributions, evolved using $\gamma$\texttt{EKO} from the GRVLO initial condition at $Q_{\rm in} = \SI{1.51}{\GeV}$, to the GRVLO parametrization as a function of $x$ at $Q^2=(16, 100, 400)$ GeV$^2$.}
\label{fig:gEKO_LO}
\end{figure}

To illustrate and validate our implementation further, we compare 
our evolution results with the GRV photon PDFs~\cite{Gluck:1991jc}. Specifically, using the GRVLO boundary condition at the initial scale $Q_\text{in} = \SI{1.51}{\GeV}$, we evolve the quark singlet and gluon distributions using $\gamma$\texttt{EKO} up to three different higher scales, $Q^2=(16, 100, 400)$ GeV$^2$. \Cref{fig:gEKO_LO} presents the ratios of the resulting quark singlet (left panel) and gluon (right panel) distributions to their reference values given by the GRVLO parametrization as a function of $x$.
One can see from the figure that for the entire range of $x$, with the exception of very large $x$, the agreement between the $\gamma$\texttt{EKO}-evolved results and the GRV parametrization is better than 2\%.

\subsection{Next-to-leading order (and beyond)}
\label{sec:gekonlo}

In the following, N$^k$LO corresponds to keeping exactly $k+1$ terms in each sum in \cref{eq:pbgk}.
Starting from NLO, the hadronic EKO $\tilde{\mathbf E}$ can no longer be obtained from a closed form expression.
Instead, a specific solution strategy for solving the DGLAP equations has to be applied (see, e.g., ref.~\cite{Candido:2022tld} for a discussion on available options).
Mathematically speaking, the formal solution of the hadronic EKO $\tilde{\mathbf E}$ is given by
\begin{align}
    \tilde{\mathbf{E}}(a_s \leftarrow a_s^0) &= \mathcal P \exp\left[-\int\limits_{a_s^0}^{a_s} \frac{\gamma(a_s')}{\beta(a_s')} \diff a_s' \right] \,,
    \label{eq:EKOdef}
\end{align}
with $\mathcal P$ the path-ordering operator for which we need to give an explicit implementation.

The so-called \enquote{truncated} strategy aims at writing the solution as a perturbative correction to the fully resummed, analytic leading-order solution, see, e.g., ref.~\cite{Vogt:2004ns} for a detailed explanation in the hadronic case.
Note that this 
method 
was also used to solve the scale evolution of photon PDFs in ref.~\cite{Gluck:1983mm}. 

In our work, we apply instead the so-called \enquote{iterative} strategy based on the algorithm outlined in ref.~\cite{Bonvini:2012sh}.
Effectively, this is similar to the solution strategy adopted by the evolution codes implementing the splitting functions in momentum fraction $x$-space.
Given a sufficiently fine-grained grid in the strong coupling, $\{a_s^k\}_0^{n+1}$, we set
\begin{equation}
    \tilde{\mathbf E}(a_s\leftarrow a_s^0) = \prod\limits_{k=n}^{0}\tilde{\mathbf E}(a_s^{k+1}\leftarrow a_s^{k}) \,,
    \label{eq:ekoiterprod}
\end{equation}
where the order of the product is such that later (partial) EKOs are to the left and the boundary condition $a_s^{n+1} = a_s$ is satisfied.
The (partial) EKOs are given by
\begin{equation}
    \tilde{\mathbf E}(a_s^{k}\leftarrow a_s^{k}) = \exp\left(-\frac{\gamma(a_s^{k+1/2})}{\beta(a_s^{k+1/2})} (a_s^{k+1}-a_s^k)\right) \,,
    \label{eq:ekoiter}
\end{equation}
which uses the half-way point $a_s^{k+1/2} = (a_s^{k+1}+a_s^k)/2$ as suggested in ref.~\cite{Bonvini:2012sh}.

We can use the same strong coupling grid, $\{a_s^k\}_0^{n+1}$, also for solving \cref{eq:inhom}, which allows us to solve the hadronic EKO $\tilde{\mathbf E}$ and the point-like contributions $\tilde{\mathbf f}_{inhom}^\gamma$ at the same time.
We split the explicit integral in \cref{eq:inhom} using the trapezoidal rule and write
\begin{align}
    \tilde{\mathbf f}^\gamma_{\rm inhom}(a_s) &= \int\limits_{a_s^0}^{a_s}\! \diff a_s'\, \tilde{\mathbf E}(a_s \leftarrow a_s') \frac{-\tilde{\mathbf k}(a_s')}{\beta(a_s')}\nonumber \\
    &\approx \sum_{k=0}^n \left( \tilde{\mathbf E}(a_s \leftarrow a_s^k) \frac{-\tilde{\mathbf k}(a_s^k)}{\beta(a_s^k)} + \tilde{\mathbf E}(a_s \leftarrow a_s^{k+1}) \frac{-\tilde{\mathbf k}(a_s^{k+1})}{\beta(a_s^{k+1})}\right) \frac{a_s^{k+1}-a_s^k}{2}\,.
    \label{eq:inhomiter}
\end{align}
By inserting \cref{eq:ekoiterprod,eq:ekoiter} into \cref{eq:inhomiter} and using the transitivity of hadronic EKOs, i.e., the explicit ordering in \cref{eq:ekoiterprod}, we can reorder the whole expressions and do in a single iteration the product of \cref{eq:ekoiterprod} and the sum of \cref{eq:inhomiter} simultaneously as required.
We find
\begin{align}
    \tilde{\mathbf f}^{\gamma,0}_{\rm inhom} &= \exp\left(-\frac{\gamma(a_s^{1/2})}{\beta(a_s^{1/2})} (a_s^{1}-a_s^0)\right)\frac{-\tilde{\mathbf k}(a_s^{0})}{\beta(a_s^{0})}\frac{a_s^1-a_s^0}{2} \,, \\
    \tilde{\mathbf f}^{\gamma,k}_{\rm inhom} &= \exp\left(-\frac{\gamma(a_s^{k+1/2})}{\beta(a_s^{k+1/2})} (a_s^{k+1}-a_s^k)\right)\left(\tilde{\mathbf f}^{\gamma,k-1}_{\rm inhom}+\frac{-\tilde{\mathbf k}(a_s^{k})}{\beta(a_s^{k})}\frac{a_s^{k+1}-a_s^{k-1}}{2}\right)\,,\quad k=1\ldots n\\
    \tilde{\mathbf f}^{\gamma,n+1}_{\rm inhom} &= \tilde{\mathbf f}^{\gamma,n}_{\rm inhom} + \frac{-\tilde{\mathbf k}(a_s^{n+1})}{\beta(a_s^{n+1})}\frac{a_s^{n+1}-a_s^n}{2} \,,
\end{align}
and, finally, we can identify $\tilde{\mathbf f}^\gamma_{\rm inhom}(a_s) = \tilde{\mathbf f}^{\gamma,n+1}_{\rm inhom}$.
This prescription works at any order and both in the non-singlet and singlet sectors (provided one uses the standard matrix exponentiation methods), where, as usual, non-commutativity is crucial.
Note that we evaluate the beta function both at the half-way point and the grid points in a single step.

\begin{figure}[t!]
    \centering
   \includegraphics[width=\linewidth]{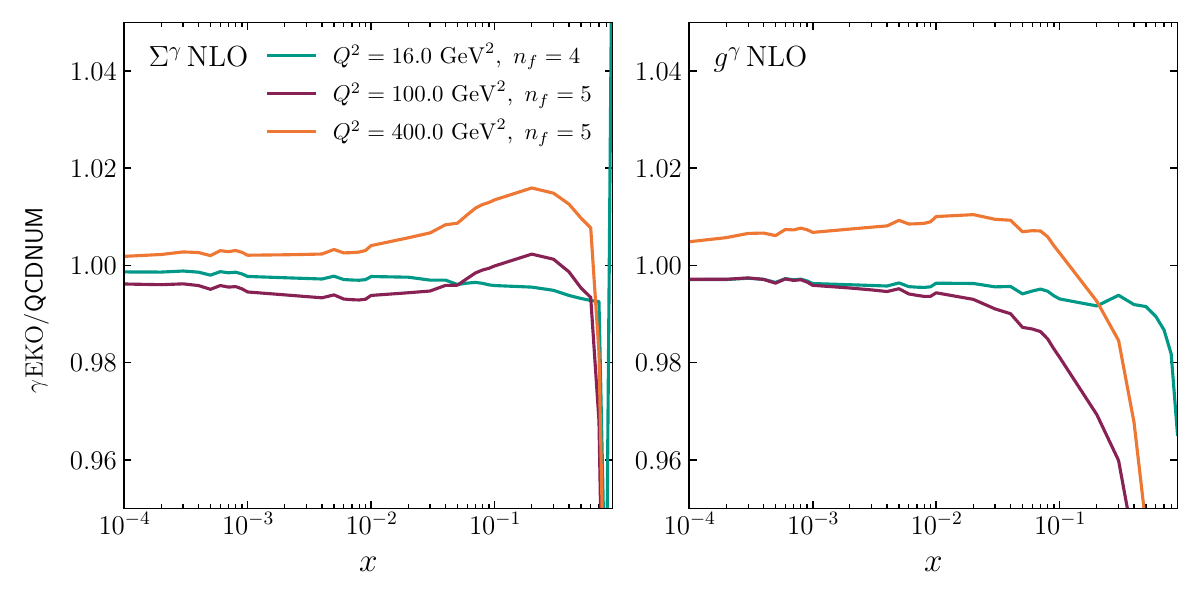}
\caption{Ratios of NLO quark singlet $\Sigma^{\gamma}(x,Q^2)$ (left panel) and gluon
$g^{\gamma}(x,Q^2)$ (right panel) distributions, evolved using $\gamma$\texttt{EKO} 
and modified \texttt{QCDNUM} from the GRVHO initial condition at $Q_{\rm in} = 1.51$ GeV, as a function of $x$ at $Q^2=(16, 100, 400)$ GeV$^2$.}
    \label{fig:gEKO_NLO}
\end{figure}

To benchmark our implementation, we compare the results of our scale evolution at NLO with reference values obtained using an appropriately modified \texttt{QCDNUM} version~\cite{Botje:2010ay}; in both cases, the boundary condition is given by the GRVHO photon PDFs at $Q_\text{in} = \SI{1.51}{\GeV}$.
\Cref{fig:gEKO_NLO} shows the ratios of the $\Sigma^{\gamma}(x,Q^2)$ and $g^{\gamma}(x,Q^2)$ distributions resulting from these two evolution methods as a function of $x$ at three values of $Q^2=(16, 100, 400)$ GeV$^2$.
One can see from the figure that the ratios of the quark singlet and gluon distributions are close to unity within a few percent, indicating a high accuracy of our numerical implementation scale evolution at NLO across all momentum fractions $x$.
Note that we opted to avoid a direct comparison with the GRVHO parametrization to minimize possible influence of interpolation errors, which may be present in the interface of the GRV PDFs.